\documentclass[twocolumn,superscriptaddress,amsmath,amssymb,pra]{revtex4-2}
\usepackage[utf8]{inputenc}
\usepackage[T1]{fontenc}
\usepackage{graphicx}
\usepackage{dcolumn}
\usepackage{appendix}
\usepackage{bm}

\usepackage[usenames,dvipsnames]{xcolor}

\usepackage{subfigure}
\usepackage{bbm}
\usepackage{enumerate}

\usepackage[
bookmarks=true,
colorlinks,
linkcolor=black,
urlcolor=blue,
citecolor=magenta,
plainpages=false,
pdfpagelabels,
final,
breaklinks=true
]{hyperref}

\usepackage{titlesec}

\usepackage{array}

\usepackage{physics}

\usepackage[left]{lineno}

\usepackage{etoolbox}
\makeatletter
\patchcmd\@outputpage{\begingroup}{\begingroup\resetlinenumber[1]}{}{}
\makeatother

\newcommand{\relaxket}[1]{\lvert{#1}\rangle}
\newcommand{\relaxbra}[1]{\langle{#1}\rvert}

\begin{document}

\title{Attosecond quantum optical interferometry }

\author{Javier Rivera-Dean}
\email{physics.jriveradean@proton.me}
\affiliation{ICFO-Institut de Ciencies Fotoniques, The Barcelona Institute of Science and Technology, Castelldefels (Barcelona) 08860, Spain}
\affiliation{Department of Physics and Astronomy, University College London, Gower Street, London WC1E 6BT, UK}

\author{Lidija Petrovic}
\affiliation{ICFO-Institut de Ciencies Fotoniques, The Barcelona Institute of Science and Technology, Castelldefels (Barcelona) 08860, Spain}

\author{Maciej Lewenstein}
\affiliation{ICFO-Institut de Ciencies Fotoniques, The Barcelona Institute of Science and Technology, Castelldefels (Barcelona) 08860, Spain.}
\affiliation{ICREA, Pg. Llu\'{\i}s Companys 23, 08010 Barcelona, Spain}

\author{Philipp Stammer}
\email{philipp.stammer@icfo.eu}
\affiliation{ICFO-Institut de Ciencies Fotoniques, The Barcelona Institute of Science and Technology, Castelldefels (Barcelona) 08860, Spain.}
\affiliation{Atominstitut, Technische Universit\"{a}t Wien, 1020 Vienna, Austria}

\date{\today}

\begin{abstract}

In this work, we explore the scheme of attosecond quantum interferometry (AQI), the quantum optical version of classical attosecond interferometry, which allows to measure quantum optical properties on the attosecond time-scale.~We develop how the scheme of AQI can be used for engineering the phase-space and photon statistics properties of the emitted harmonics, using the relative phase of a two-color driving field as a control, and further enables to manipulate the field correlations as well as their entanglement characteristics.~In addition, this scheme allows us to learn properties of the phase-space distribution of the harmonic quantum state, by means of measuring an attosecond quantum tomography trace.~This serves as a new type of protocol for \emph{in situ} attosecond measurements of quantum optical observables.~With this, we achieve to further connect all-optical attosecond measurement schemes with quantum optics, allowing for a rich manifold of observations.

\end{abstract}

\maketitle

\section{\label{sec:intro}INTRODUCTION}

Quantum optics of intense laser-driven processes has experienced a rapid development over the recent years~\cite{cruz2024quantum, stammer2025colloquium, lewenstein2021generation}, providing novel insights into the up-conversion process of high harmonics generation (HHG), challenging the folklore of previous wisdom~\cite{rivera2025structured}.~While HHG was successfully described by semi-classical methods for decades~\cite{corkum1993plasma, lewenstein1994theory, amini2019symphony}, recent achievements have shown that non-trivial systems can lead to squeezing signatures in the emitted harmonics~\cite{lange2024electron, lange2025excitonic, yi2024generation,theidel_observation_2025}, that the harmonic field modes can be entangled~\cite{stammer2022high, stammer2024entanglement, theidel2024evidence, yi2024generation} or that the emitted photons can have anti-bunching statistics~\cite{stammer2025theory}.

However, quantum optical HHG still faces the challenge of ambiguities about the presence of genuine quantum signatures, as well as the quest for clear applications~\cite{stammer2025colloquium}. For instance, the quantum state of the emitted harmonic radiation from a correlated system can show deviations from Gaussian coherent states while still being completely classical~\cite{pizzi2023light}, while in contrast, the state can be Gaussian but genuine quantum due to squeezing~\cite{stammer2024entanglement}. 
Furthermore, driving the process of HHG with quantum light~\cite{rasputnyi2024high}, such as bright squeezed vacuum, revealed new insights, for instance an extended cutoff in the harmonic spectrum~\cite{gorlach2023high}.~However, despite being driven by quantum light, the extended cutoff remains a purely classical signature reproducible by classical thermal driving fields~\cite{gorlach2023high}. 
In addition, driving HHG by a combination of classical coherent and squeezed light has been shown very recently to generate quantum states reminiscent of squeezed states~\cite{tzur2025measuring, stammer_weak_2025}. Nevertheless, these states do likewise not show genuine quantum properties below shot noise squeezing~\cite{loudon1987squeezed}, and therefore remain classical. 

\begin{figure}
	\centering
	\includegraphics[width=0.9\columnwidth]{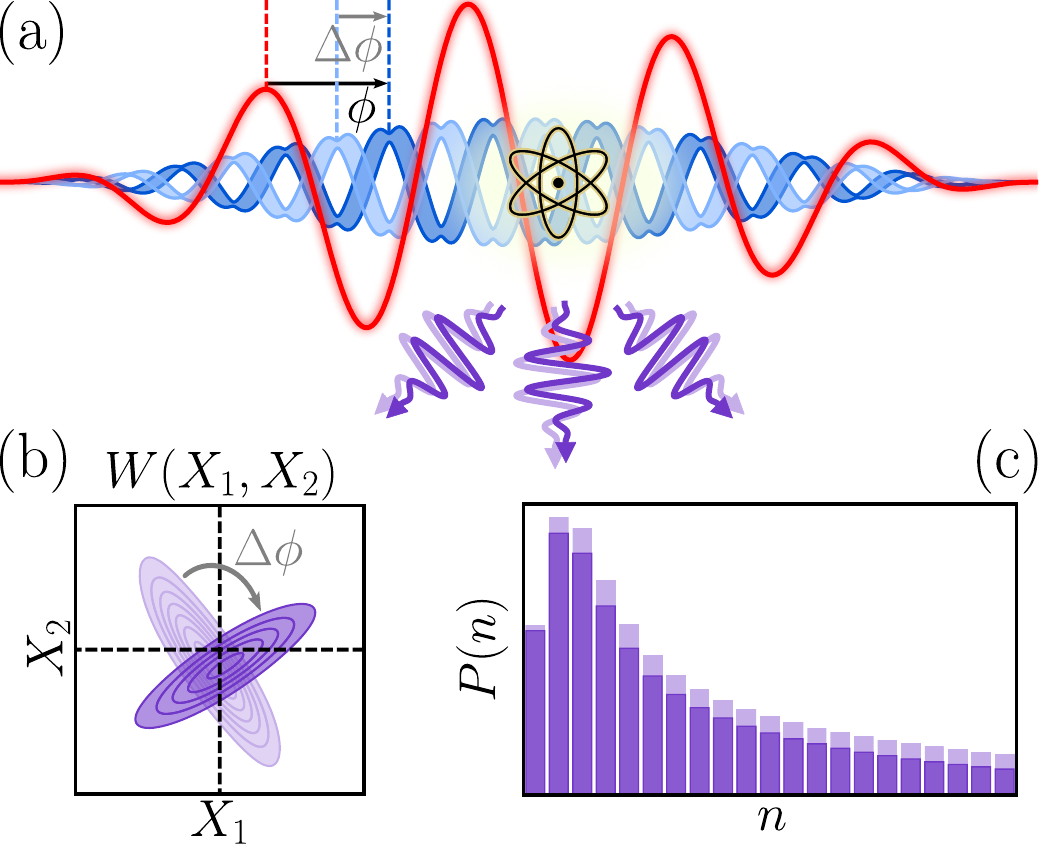}
	\caption{\textbf{Attosecond quantum interferometry.}~(a) Pictorial representation of the scheme under consideration.~A strong classical pump field at frequency $\omega$ is combined with a perturbative $2\omega$ field exhibiting amplitude squeezing.~By varying the relative phase $\phi$ between the two fields, one can (b) control the phase-space distribution and orientation of the generated harmonics, displayed here in terms of their Wigner function; (c) modify their photon-number statistics, peaked around the mean photon number obtained in the absence of squeezing; as well as tailor its intermode correlations.~For representational purposes, the harmonics are shown as emitted perpendicular to the polarization direction; in practice, emission occurs along the laser propagation direction. }
	\label{fig:scheme}
\end{figure}

The aforementioned configurations of two-color driving fields, with one component being perturbative, have been explored in the classical setting with two coherent driving fields, and are usually termed attosecond interferometry~\cite{dahlstrom2011quantum, pedatzur2015attosecond, dudovich2006measuring, heck2021attosecond, brown2022attosecond}. Such all-optical interferometry schemes rely on the interference from the electron dynamics in different half-cycles of the driving pulse, where the presence of a perturbative second harmonic field breaks the symmetry between these events, leading to the presence of even harmonic orders.~This, in turn, enables \emph{in situ} measurement of attosecond pulses~\cite{dudovich2006measuring}, as well as the reconstruction of information on the strong-field driven electron dynamics with ultrafast resolution~\cite{dahlstrom2011quantum, pedatzur2015attosecond}.

In this work, we consider its fully quantized counterpart, a concept introduced as \textit{attosecond quantum interferometry (AQI)}~\cite{stammer_weak_2025}. We focus on a configuration consisting of a strong coherent classical pump at frequency $\omega$ combined with a perturbative second harmonic $2 \omega$ prepared in a squeezed quantum field (see Fig.~\ref{fig:scheme}~(a) for a schematic configuration). After establishing the theoretical framework that sets the foundation of our analysis (Sec.~\ref{Sec:Th:Back}), we show how AQI enables phase-sensitive control over the phase-space properties of the harmonic field modes.~In particular, varying the relative phase between the two colors effectively modifies the Wigner function of the harmonics and allows for a controllable rotation in phase space~[Fig.~\ref{fig:scheme}~(b)], while changing their statistical properties~(Sec.~\ref{sec:2}). We then demonstrate that this control extends beyond the photon statistics of isolated harmonics~[Fig.~\ref{fig:scheme}~(c)] to the correlations between them, with characteristics that depend strongly on the even or odd order of the generated harmonics~(Sec.~\ref{Sec:Field:corr}). Building on this phase-dependent control mechanism, we introduce a method to access information about the harmonics phase-space distribution in an otherwise experimentally inaccessible spectral regime via an \emph{attosecond quantum tomography (AQT)} protocol, which we dub the AQT-distribution~(Sec.~\ref{Sec:AQT}). Since the same phase parameter that controls the rotation of the harmonic phase-space distribution also alters the harmonic statistics itself, we critically assess the scope, capabilities and limitations of AQT as a quantum tomography-like technique.

\section{\label{Sec:Th:Back}THEORETICAL BACKGROUND}
Leveraging the approaches from classical attosecond interferometry experiments~\cite{dahlstrom2011quantum, pedatzur2015attosecond}, where an intense driving field and its perturbative second harmonic induce the process of HHG, we consider the quantum optical version of this scheme and extend the analysis by including non-classical driving fields~\cite{rivera2025structured, stammer_weak_2025}.~This is achieved by substituting the classical perturbative second harmonic with a squeezed state.~The total driving field is accordingly given by 
\begin{equation}\label{Eq:init:state}
	\ket{\Phi_{\text{drive}}(t_0)}
	= \ket{\alpha_{\omega}}
	\otimes \hat{D}(\alpha_{2\omega}) \hat{S}(\xi) \ket{0},
\end{equation}
where the second harmonic amplitude $\alpha_{2\omega} = \abs{\alpha_{2\omega}} e^{i\phi}$, is perturbative $\abs{\alpha_{2\omega}} = \epsilon \abs{\alpha_{\omega}}$ with $\epsilon \ll 1$.~Here, we consider the scenario in which the squeezing parameter is given by $\xi(\phi) = r e^{i\varphi(\phi)}$, with fixed amplitude $r$ and a phase $\varphi$ that varies with the two-color delay $\phi$ according to $\varphi = 2\phi$.~In this way, the $2\omega$ field remains amplitude squeezed independently of the tunable phase difference $\phi$~\cite{stammer_weak_2025}.~The squeezing intensity is fixed to $I_{\text{squ}} = \kappa^2 \sinh[2](\abs{\xi}) = 10^{-6}$ a.u., a value compatible with recent experimental implementations in the strong-field regime~\cite{lemieux2024photon,rasputnyi2024high,tzur2025measuring}.~Here, $\kappa = \sqrt{\hbar\omega/(2\epsilon_0 V)}$ denotes the light-matter coupling, with $V$ the quantization volume~\cite{lewenstein2021generation, rivera2025structured}.  

\subsection{Light-matter interaction dynamics}
An equivalent density matrix representation of the initial driving field state in Eq.~\eqref{Eq:init:state} can be cast in terms of the generalized positive-$P$ representation~\cite{drummond1980generalised}
\begin{equation}
	\hat{\rho}_{\text{drive}}(t_0)
		= \dyad{\alpha_\omega}\otimes 
			\int\dd^2 \alpha \int \dd^2 \beta
				\dfrac{P_{2\omega}(\alpha,\beta^*)}{\braket{\beta^*}{\alpha}}
					\dyad{\alpha}{\beta^*},
\end{equation}
where $P_{2\omega}(\alpha,\beta^*)$ is a positive-definite distribution encoding the quantum properties of the $2\omega$ driver. At any time $t\geq t_0$, the total quantum state of the system (including the emitter and harmonics) evolves according to the von Neumann equation
\begin{equation}\label{Eq:vNdiff}
	i\hbar \pdv{\hat{\rho}(t)}{t}
		= \big[
				\hat{H},\hat{\rho}(t)
			\big],
\end{equation}
with initial condition $\hat{\rho}(t_0) = \dyad{\text{g}}\otimes \hat{\rho}_{\text{drive}}(t_0)\otimes \dyad{\bar{0}}$. Here, $\ket{\text{g}}$ denotes the atomic ground state and $\ket{\bar{0}} \equiv \bigotimes_{q>2}\dyad{0_{q\omega}}$ represents the state of all harmonic modes beyond the fundamental and its second harmonics.~The total Hamiltonian is given by $\hat{H} = \hat{H}_{\text{at}} + \mathsf{e}\hat{r} \hat{E}(t) + \hat{H}_{\text{field}}$, where $\hat{H}_{\text{at}}$ is the atomic Hamiltonian, and the (linearly polarized) electric field-operator reads $\hat{E} = -i \kappa \sum_{q}\sqrt{q}[\hat{a}_{q}-\hat{a}_{q}^\dagger]$. The free-field Hamiltonian is $\hat{H}_{\text{field}} = \sum_{q}\hbar q\omega \hat{a}^\dagger_{q}\hat{a}_{q}$.

Under the strong-field approximation~\cite{lewenstein1994theory,lewenstein2021generation} and low-depletion conditions~\cite{stammer2024entanglement}, the joint light-matter state at a final time $t$ can be approximately written as~\cite{gorlach2023high}
\begin{equation}\label{Eq:joint:state}
	\begin{aligned}
		\hat{\rho}(t)
		\approx \int \dd^2\alpha \int \dd^2 \beta&\
		\dfrac{P(\alpha,\beta^*)}{\braket{\beta^*}{\alpha}}
		\dyad{\psi_\alpha(t)}{\psi_{\beta^*}(t)}
		\\&
		\otimes
		\dyad{\Phi_{\alpha}(t)}{\Phi_{\beta^*}(t)},
	\end{aligned}
\end{equation}
where $\ket{\psi_\alpha(t)}$ denotes the electronic state at time $t$, obtained from the semi-classical dynamics
\begin{equation}
	i\hbar\pdv{\ket{\psi_{\alpha}(t)}}{t}
	=\big[
	\hat{H}_{\text{at}} + \mathsf{e} \hat{\vb{r}}\cdot \boldsymbol{E}_{\alpha}(t)        
	\big] \ket{\psi_{\alpha}(t)}, 
\end{equation}
driven by the intense classical electric field $\boldsymbol{E}_{\alpha}(t) = \text{tr}[\dyad{\alpha_\omega,\alpha,\bar{0}} \hat{\boldsymbol{E}}(t)]$, where $\hat{\boldsymbol{E}}(t) = e^{i\hat{H}_{\text{field}}t/\hbar}\hat{\boldsymbol{E}}e^{-i\hat{H}_{\text{field}}t/\hbar}$. The state of the field $\ket{\Phi_{\alpha}(t)}$ is obtained from the conventional quantum optical solution to HHG in the low-depletion regime~\cite{lewenstein2021generation,rivera2022strong,stammer2023quantum}
\begin{equation}
	\label{Eq:solution}
	\ket{\Phi_{\alpha}(t)}
	= \big[
	\hat{D}_{\omega}(\alpha_\omega)\otimes\hat{D}_{2\omega}(\alpha)
	\big] \bigotimes_{q=1} \ket{\chi_{q,\alpha}(t)},
\end{equation}
where $\chi_q(t) = \int^{t}_{t_0}\dd \tau \langle \psi_{\alpha}(\tau)\vert \hat{r}\vert \psi_{\alpha}(\tau) \rangle e^{i\omega_q t}$ are the coherent amplitudes due to the field induced charge current of the electron. It is worth noting that, although the above solution has been derived within the single-active electron approximation, in practice the strong-field response of a gaseous medium involves $N$ uncorrelated emitters. In such a case, owing to the absence of interatomic correlations, the total harmonic response is obtained as the sum of the independent single-atom contributions. Consequently, the harmonic amplitude scales linearly with the number of emitters, yielding an effective rescaling $\chi_{q,\alpha}(t) \to N\chi_{q,\alpha}(t)$~\cite{stammer2023quantum,stammer2025theory}. In the following, we explicitly incorporate this scaling factor into our expressions.

Finally, since we are interested in the state and the properties of the emitted harmonics, we can use that the final quantum optical state associated with the $q$-th harmonic mode, after tracing out the both the driving field and electronic degrees of freedom, can be expressed as~\cite{even_tzur_photon-statistics_2023}
\begin{equation}\label{Eq:Meth:state}
	\begin{aligned}
		\hat{\rho}_q(t)
		= \int \dd^2\alpha \int \dd^2 \beta&\
		\dfrac{P(\alpha,\beta^*)}{\langle N\chi^{(q)}_{\beta^*}(t)\vert N\chi^{(q)}_{\alpha}(t)\rangle}
		\\&\hspace{1cm}\times	\relaxket{N\chi^{(q)}_{\alpha}(t)}\!\relaxbra{N\chi^{(q)}_{\beta^*}(t)}.
	\end{aligned}
\end{equation}

\subsection{Evaluation of harmonic observables}
Having solved for the quantum state of the harmonics, we can now investigate its properties and how to control them via the two-color phase difference $\phi$.~To do so, and considering a general quantum optical observable $\hat{O}_q$ of the $q$-th harmonic mode, it follows from Eq.~\eqref{Eq:Meth:state} that its expected value reads
\begin{equation}\label{Eq:Meth:Obs}
	\begin{aligned}
		\langle\hat{O}_q\rangle
		= \int \dd^2\alpha \int \dd^2 \beta&\
		\dfrac{P(\alpha,\beta^*)}{\langle N\chi^{(q)}_{\beta^*}(t)\vert N\chi^{(q)}_{\alpha}(t)\rangle}
		\\&\hspace{0.5cm}\times
		\langle N \chi^{(q)}_{\alpha}(t)
		\vert\hat{O}_q\vert N\chi^{(q)}_{\beta^*}(t)\rangle.
	\end{aligned}
\end{equation} 

Using that the driving fields consist of many photons, it is convenient to evaluate the corresponding observables in the joint classical and quasi-thermodynamic limit~\cite{gorlach2023high,stammer_weak_2025} (see Supplementary Material). This regime is natural for strong-field scenarios, where the interaction takes place in free space, with high field intensity, and for a macroscopic number of emitters.~Formally, we take $(V,\alpha,N) \to \infty$ while keeping the field strength $\varepsilon_\alpha = 2\kappa \alpha$ and $\varrho = N \kappa$ constant. Under these conditions, Eq.~\eqref{Eq:Meth:Obs} reduces to (see Supplementary Material)
\begin{equation}\label{Eq:obs:after:limit}
	\langle\hat{O}_q\rangle
	= 
	\int \dd \varepsilon_\alpha
	\mathcal{Q}(\varepsilon_\alpha)
	\langle \varrho \, d_{\varepsilon_{\alpha}}(\omega_q)\vert \hat{O}_q\vert \varrho \, d_{\varepsilon_{\alpha}}(\omega_q)\rangle,
\end{equation}
where $\mathcal{Q}(\varepsilon_\alpha)$ is the Husimi $Q$-function evaluated in the classical and quasi-thermodynamic limits.

Equation~\eqref{Eq:obs:after:limit} highlights a crucial feature of the quantum-optical observables of individual harmonics:~their behavior is effectively classical, since the averages are obtained from the corresponding statistical mixture
\begin{equation}
    \hat{{\rho}}_q = \int \dd \varepsilon_{\alpha} \mathcal{Q}(\varepsilon_\alpha)\dyad{\varrho \, d_{\varepsilon_{\alpha}}(\omega_q)}.
\end{equation}
While the quasi-thermodynamic and classical limits leading to this result removes coherences of the form $\lvert\chi^{(q)}_{\alpha}(t)\rangle\!\langle\chi^{(q)}_{\beta^*}(t)\vert$, the same conclusions arises in analytical treatments that do not rely on such limits~\cite{wang_high-order_2025}.~This reflects the intrinsic physics of the high-photon number regime.~Even without the aforementioned limits, the off-diagonal contributions from the coherences are weighted by $\text{exp}[-\abs{\alpha - \beta^*}^2/4]$, which decays exponentially with the distance between $\alpha$ and $\beta^*$.~At high intensities, where hundreds or even thousands of photons are involved, small differences between $\alpha$ and $\beta^*$ have a negligible influence on the phase or amplitude of $\chi_{\alpha}^{(q)}(t)$ and $\chi^{(q)}_{\beta^*}(t)$, but they strongly suppress the off-diagonal elements in $P(\alpha,\beta^*)$. This exponential suppression ensures that the state is effectively a classical diagonal distribution of coherent states, justifying the form of Eq.~\eqref{Eq:obs:after:limit}. An alternative derivation of this result is presented in the Supplementary Material, where the dynamics are solved using an expansion on the coherent-state basis of the initial field, rather than employing the generalized positive-$P$ representation. 

\section{\label{sec:1}RESULTS}

Despite the fact that the expectation values of quantum optical observables for the harmonics admit an effective classical description, the underlying quantum states remain highly nontrivial.~The following analysis explores the phase-space structure of the generated harmonic states, highlighting how their photon statistics and intermodal correlations can be controlled.~In this direction, the two-color delay $\phi$ emerges as a tunable control parameter, providing insight into the nature of the driving fields while simultaneously enabling the extraction of tomographic information through AQT.~We examine the scope and limitations of AQT as a genuine quantum-tomography-like protocol, and benchmark its performance against exact homodyne tomography.

\subsection{\label{sec:2}Control of the harmonic quantum state}
To characterize the phase-space structure of the harmonic quantum states, we evaluate their Wigner function, which provides a direct representation of the quantum state in phase-space~\cite{wigner_quantum_1932}.~The Wigner function constitutes a complete and equivalent description of the state, fully consistent with other standard formulations such as the density matrix representation~\cite{SchleichBookCh3}, and therefore encodes all of its information.~Importantly, established quantum optical protocols such as homodyne detection enable their experimental reconstruction~\cite{smithey_measurement_1993}, making the Wigner function a particularly valuable tool for doing a tomographic reconstruction of quantum optical states.~Theoretically, for an arbitrary quantum state $\hat{\rho}$, its Wigner function can be expressed as~\cite{royer_wigner_1977}
\begin{equation}\label{Eq:Wigner:function}
	W(X_1,X_2)
		= \text{tr}[\hat{\mathcal{W}} \hat{\rho}],\ 
			\text{with}\  
			\hat{\mathcal{W}} = \hat{D}(\beta)\hat{\Pi}\hat{D}^\dagger(\beta),
\end{equation}
where $\hat{\Pi}$ denotes the parity operator, and the optical quadratures are defined as $X_1 \equiv \text{Re}[\beta]$ and $X_2 \equiv \text{Im}[\beta]$. 

A comparison between Eqs.~\eqref{Eq:obs:after:limit} and \eqref{Eq:Wigner:function} shows that, under the strong-field conditions considered here, the harmonics Wigner function reduce to a $\mathcal{Q}$-weighted statistical mixture of coherent state Wigner functions.~Since coherent states possess strictly positive Wigner distributions~\cite{hudson_when_1974}, the resulting harmonic Wigner functions are therefore positive-definite.~Nevertheless, we find that these exhibit a nontrivial phase-space structure that depends sensitively on both the parity of the harmonic order and the two-color delay $\phi$, and encode valuable information about the nature of the driving field.

\begin{figure}
	\centering
	\includegraphics[width=1\columnwidth]{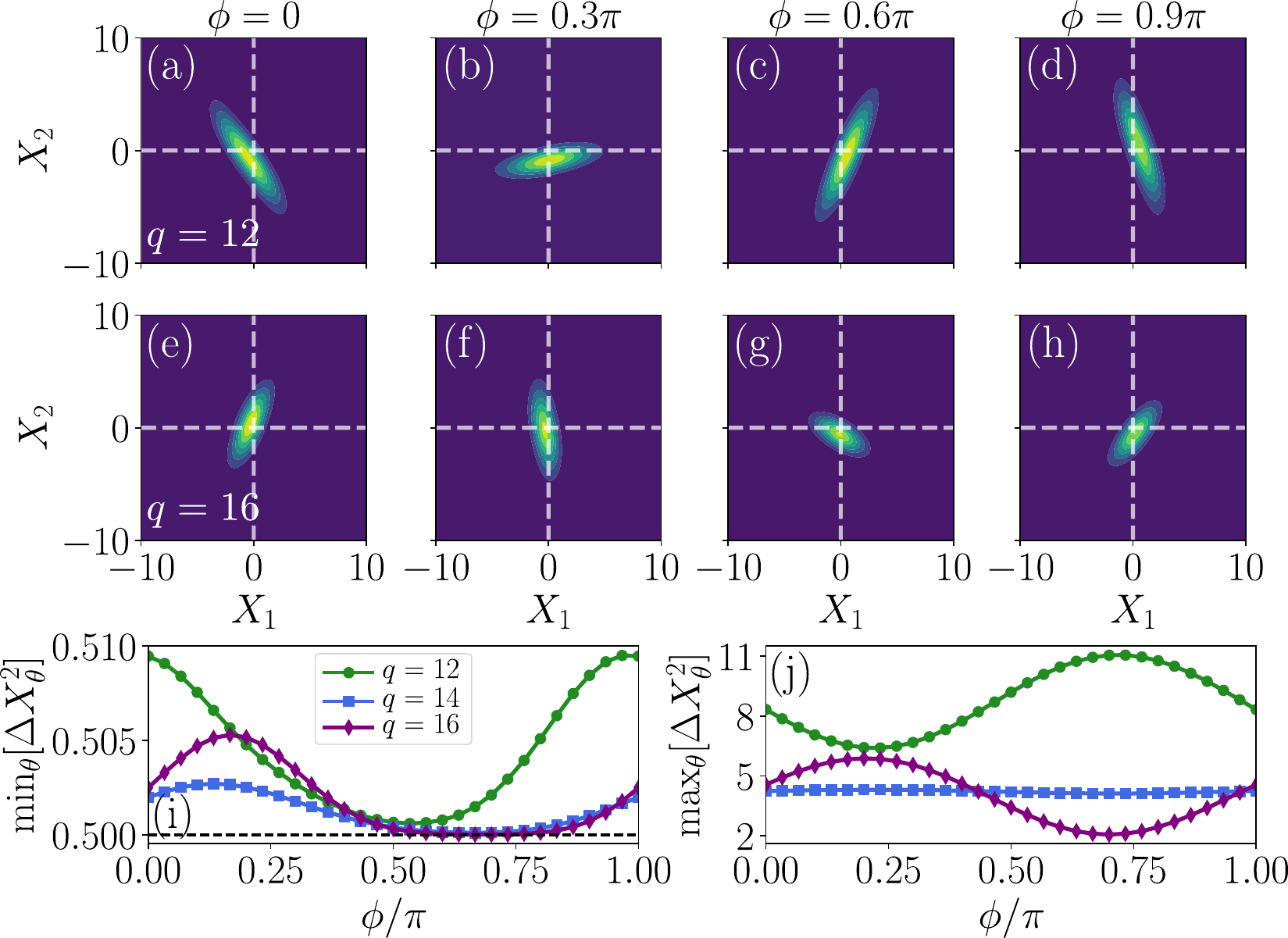}
	\caption{\textbf{Phase-space properties of the harmonics.}~Wigner functions of the harmonic modes $q=12$ [(a)-(d)] and $q=16$ [(e)-(h)] for varying two-color phase $\phi$.~Minimum [(i)] and maximum [(j)] values of the quadrature variances for three even harmonic orders.~Calculations were performed with $E_{\omega} = 0.053$ a.u., $E_{2\omega} = 10^{-2}E_{\omega}$, $I_{\text{squ}} = 10^{-6}$ a.u., $\omega = 0.057$ a.u. and $I_p =0.5$ a.u., with a field duration of 5 optical cycles. Amplitude squeezing for the $2\omega$ field is considered here although results are not affected much by the specific type of squeezing, although significantly enhanced when $I_{\text{squ}}$ increases (see Supplementary Material).}
	\label{fig:state_engineering}
\end{figure}

Figure~\ref{fig:state_engineering}~(a)-(h) displays the Wigner functions of the even harmonics $q=12$ and $q=16$ for varying relative phase $\phi$. In both cases, features reminiscent of quadrature squeezing are observed, with a strong dependence on the harmonic order, and with the squeezing orientation rotating counter-clockwise as $\phi$ goes from 0 to $2\pi$.~Conversely, odd harmonics exhibit classical-like coherent Wigner functions~\cite{stammer_weak_2025}, due to their origin in the strong $\omega$ field, largely unaffected by the weaker $2\omega$ component.~Since even harmonics, by contrast, only appear when adding the $2\omega$ field, they inherit the properties of the perturbation: both their strength and shape are dependent on how the $2\omega$ field perturbs the $\omega$-driven ionization dynamics~\cite{pedatzur2015attosecond}.

To quantify the apparent squeezing-like features, we evaluate the minimal and maximal quadrature variances, $\operatorname{min}_\theta [(\Delta X_\theta)^2]$ and $\operatorname{max}_\theta [(\Delta X_\theta)^2]$, with $\hat{X}_{\theta} = \hat{a}^\dagger e^{i\theta} + \hat{a}e^{-i\theta}$, the results shown in Fig.~\ref{fig:state_engineering}~(i) and (j), respectively, for three different harmonic orders.~While the maximal variance lies well above the shot-noise level of $(\Delta X_{\theta})^2 = 0.5$, the minimal variance never falls below the vacuum limit, but remains particularly close to it, reaching values of  $\operatorname{min}_\theta [(\Delta X_\theta)^2] \approx 0.51$ at most.~This confirms that no genuine quantum squeezing is present: one quadrature is classically stretched, and the conjugate displays vacuum noise correlations, with the orientation of this anisotropy depending strongly on the two-color phase delay. This $\phi$-dependence originates from the squeezed nature of the $2\omega$ driving field, as the phase difference between the two colors determines the squeezed quadrature of the $2\omega$ component with respect to the quadrature reference set by the coherent $\omega$ field.~This, in turn, determines both the mean intensity of the generated harmonics and their intensity fluctuations~\cite{pedatzur2015attosecond} .

\begin{figure}
	\centering
	\includegraphics[width=1\columnwidth]{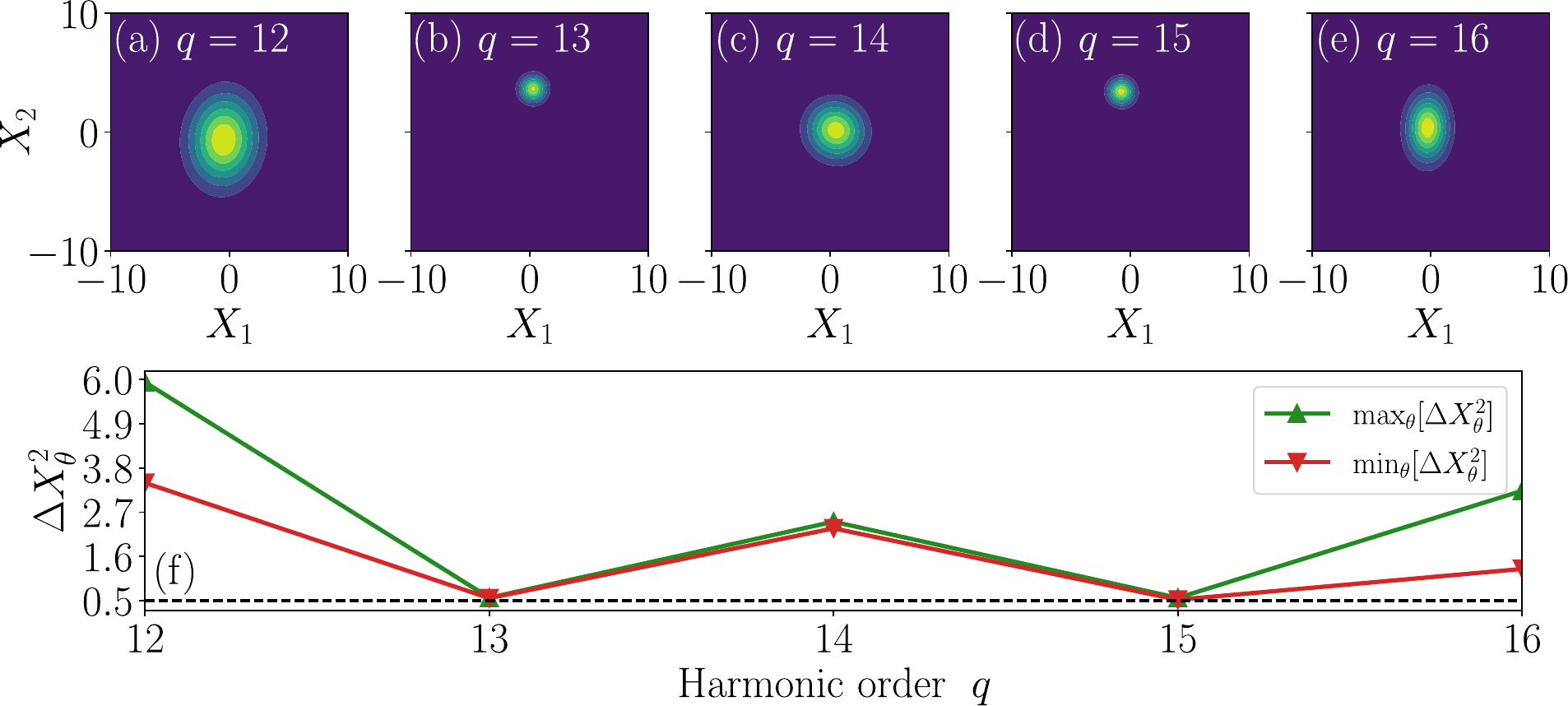}
	\caption{\textbf{Analysis for a $\boldsymbol{2\omega}$ thermal state driver.}~(a)-(e) Wigner function of various harmonic orders for a driving field composed of a strong coherent $\omega$ component and a displaced thermal state at $2\omega$.~(f) Maximum (green curve) and minimum (red curve) quadrature variances for the same harmonic orders.~Calculations were performed with $E_{\omega} = 0.053$ a.u., $E_{2\omega} = 10^{-2}E_{\omega}$, $I_{\text{th}} = 10^{-6}$ a.u., $\omega = 0.057$ a.u. and $I_p =0.5$ a.u., with a field duration of 5 optical cycles.}
	\label{Fig:thermal}
\end{figure}

Interestingly, the sensitivity of the harmonic radiation to the relative squeezing orientation suggests that the $\omega-2\omega$ configuration could serve as a diagnostic for identifying squeezing features in the driving field.~To illustrate this point, we consider the case where the $2\omega$ field is prepared in a thermal state which, unlike squeezed states, exhibit isotropic fluctuations across all optical quadratures and are therefore rotationally invariant in phase space.~In this scenario, and as shown in Fig.~\ref{Fig:thermal}~(a)-(e), the even harmonic orders display Wigner functions with increased quadrature flutuations that are homogeneous in phase space compared to the squeezed-driving case.~Odd harmonics, instead, maintain their coherent-like behavior.~This is reflected quantitatively in Fig.~\ref{Fig:thermal}~(f), where both the minimal and maximal quadrature variances exhibit a substantial excess of noise above the vacuum level for the even harmonics, while reaching the vacuum level for the odd modes.~Consequently, a pronounced increase in $\text{min}_{\theta}[(\Delta X_{\theta})^2]$ could serve as an indirect, yet sensitive, witness of squeezed-light characteristics in the driver, even in the high-photon regime where direct quantum-optical measurements are challenging.

\subsection{Photon statistics and field correlations}\label{Sec:Field:corr}
Given that adding squeezing features to the $2\omega$ field and varying its relative phase with respect to the $\omega$ driver quantitatively changes the field fluctuations of the emitted harmonics, a natural question arises:~can squeezed drivers also control the photon statistics of individual harmonics and the correlations between them as a function of $\phi$? To address this question, we analyze second-order field correlation functions, which provide a compact characterization of both photon statistics and intermodal correlations.~For two harmonic orders $q_1$ and $q_2$, the normalized second-order correlation function is defined as
\begin{equation}
	g_{q_1,q_2}^{(2)}
		= \dfrac{\langle \hat{a}_{q_1}^\dagger \hat{a}_{q_2}^\dagger \hat{a}_{q_1}\hat{a}_{q_2}\rangle}{\langle \hat{a}_{q_2}^\dagger \hat{a}_{q_2}\rangle \langle\hat{a}_{q_1}^\dagger \hat{a}_{q_1}\rangle }.
\end{equation}
In the absence of correlations between distinct modes ($q_1 \neq q_2$), one obtains $g_{q_1,q_2}^{(2)}=1$.~The case $q_1=q_2$ corresponds to the  autocorrelation function $g^{(2)}_q$, whose value characterizes the photon statistics of the harmonic mode: sub-Poissonian ($g_{q}^{(2)} < 1$), Poissonian ($g_{q}^{(2)} = 1$) or super-Possonian ($g_{q}^{(2)} > 1$).

In the absence of squeezing, and within the low-depletion regime considered here, the post-HHG state factorizes into the product of coherent states~\cite{stammer2024entanglement,stammer_weak_2025}, such that the harmonics are uncorrelated with each other. In this case, the second-order field correlation functions factorize, and reduce to $g^{(2)}_{q_1,q_2}  = 1$ for any pair $(q_1,q_2)$. More generally, these correlations satisfy the Cauchy-Schwarz inequality (CSI)
\begin{align}
	\left[ g^{(2)}_{q_1, q_2} \right]^2
	\leq g^{(2)}_{q_1, q_1} g^{(2)}_{q_2, q_2}, 
\end{align}
whose violation constitutes a sufficient but not necessary signature of entanglement between the harmonic modes $q_1$ and $q_2$~\cite{theidel2024evidence}. 

\begin{figure}
	\centering
	\includegraphics[width=1\columnwidth]{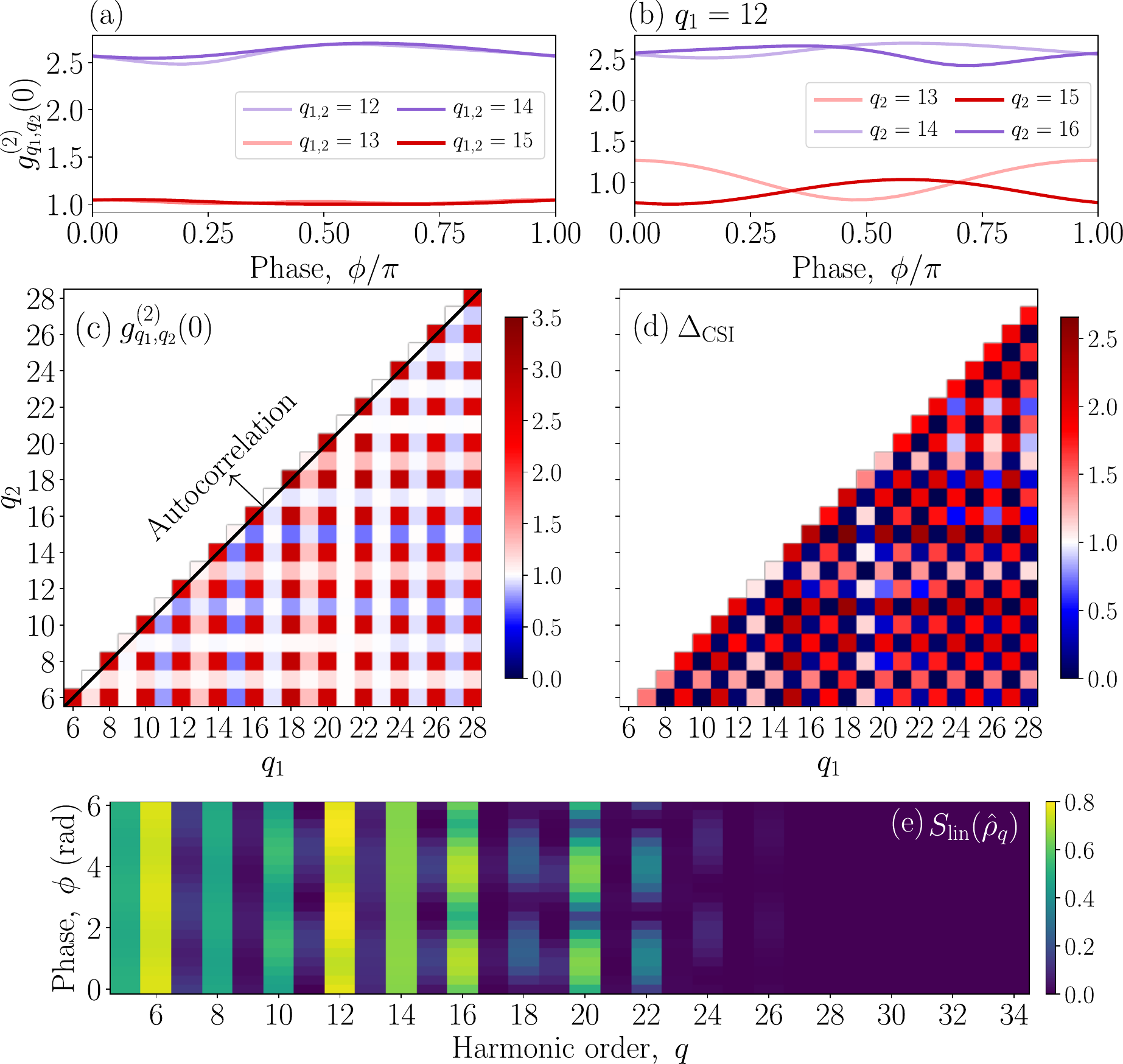}
	\caption{\textbf{Photon statistics and field correlations.} (a) Second-order autocorrelation function for high harmonic orders as a function of the two-color phase $\phi$.~(b) Second-order cross-correlation function $g^{(2)}_{12,q}$ between the 12th harmonic and other harmonic orders as a function of $\phi$.~(c) Full second-order correlation map $g^{(2)}_{q_1,q_2}$ between harmonic orders $q_1$ and $q_2$, with the black line quantifying the autocorrelations ($q_1 = q_2$), for a fixed $\phi$. (d) Value of the CSI difference. (e) Linear entropy as a function of $\phi$. The same parameters as those in Fig.~\ref{fig:state_engineering} have been considered here.}
	\label{fig:correlations}
\end{figure}

Figure~\ref{fig:correlations}~(a) and (b) display the autocorrelation for selected harmonics orders and the cross-correlation between the 12th harmonic and other orders, respectively, as a function of the two-color phase $\phi$.~As seen in panel (a), odd harmonics exhibit almost Poissonian statistics ($g^{(2)}_{\text{odd}} \approx 1$), with only a weak modulation induced by the pertuabtive field.~In contrast, even harmonics display clear super-Poissonian behavior with $g^{(2)}_{\text{even}} >2$.~These correlations are more strongly modulated by $\phi$, varying approximately between 2.5 and 2.8, reflecting both their sensitivity to intensity fluctuations of the $2\omega$ component and the relative orientation of its squeezing quadrature with respect to the coherent $\omega$ field.~Furthermore, the cross-correlations, exhibit an even stronger sensitivity to the two-color phase, highlighting a central aspect already hinted at in Fig.~\ref{fig:state_engineering}:~the two-color delay not only controls the phase-space orientation of the harmonic states but also governs their photon-number statistics and inter-harmonic correlations.

A broader view of how these correlations are distributed across harmonic orders for a fixed value of $\phi$ is presented in Fig.~\ref{fig:correlations}~(c), with the black diagonal line marking the autocorrelations. For the cross-correlations $q_1\neq q_2$, we find that even harmonics are positively correlated with each other, consistent with their common dependence on the $2\omega$ driving intensity and fluctuations.~By contrast their correlation with odd harmonic orders is almost absent, $g^{(2)}_{q_1, q_2} \approx 1$, which is expected since the intensity of the odd harmonic orders is barely influenced by the $2\omega$ contribution which only dictates the behavior of the even orders.~Importantly, no violation of the CSI is observed, indicating that the intensity correlations are consistent with a classical description of the harmonic fields state.~This is confirmed in Fig.~\ref{fig:correlations}~(d), which displays the CSI difference
\begin{equation}
	\Delta_{\text{CSI}} = g^{(2)}_{q_1, q_1} g^{(2)}_{q_2, q_2} - [g^{(2)}_{q_1, q_2}]^2.
\end{equation}
In all cases, the CSI is obeyed as $\Delta_{\text{CSI}} \geq 0$, with the equality bound $\Delta_{\text{CSI}}=0$ nearly saturated for correlations within the even or odd harmonic subspace.~In contrast, the CSI difference between the cross-correlations of even and odd harmonics is significantly larger, leading to the checkerboard-like pattern.~This is compatible with the fact that, similarly to the evaluation of single-mode harmonic observables given in Eq.~\eqref{Eq:obs:after:limit}, two-mode harmonic observables $\hat{O}_{q_1,q_2}$ can, under the same assumptions, be expressed as
\begin{equation}\label{Eq:2modeobs:after:limit}
	\begin{aligned}
	&\langle \hat{O}_{q_1,q_2}\rangle
		\\&\quad=  
		\int \dd \varepsilon_\alpha
			\mathcal{Q}(\varepsilon_\alpha)
				\tr[\hat{O}_{q_1,q_2}  \hat{\sigma}_{q_1}(\varepsilon_\alpha)\otimes \hat{\sigma}_{q_2}(\varepsilon_\alpha)],
	\end{aligned}
\end{equation}
where $\sigma_q(\varepsilon) = \dyad{\varrho d_{\varepsilon}(\omega_q)}$.

This representation shows that the correlations observed between any pair of harmonic modes are fully compatible with those of the classically correlated mixed state $\int \dd \varepsilon \mathcal{Q}(\varepsilon_\alpha) \hat{\sigma}_{q_1}(\varepsilon_\alpha)\otimes \hat{\sigma}_{q_2}(\varepsilon_\alpha)$. However, this does not exclude the presence of entanglement beyond pair-wise harmonic modes, and we therefore approach the question of field mode entanglement from a different perspective. In fact, the joint light-matter system evolves unitarily in the absence of an external environment, implying that a pure initial state must remain pure throughout the evolution.~Nonetheless, as shown by Eqs.~\eqref{Eq:obs:after:limit} and Eq.~\eqref{Eq:2modeobs:after:limit}, quantum optical expectation values of the emitted harmonics are obtained from an effective statistical mixture once all other degrees of freedom have been traced out.~This indicates that, upon tracing out all but the $q$-th harmonic mode, the quantum correlations with the remaining modes manifest as a reduced purity of the single-mode state.

In this context, the remaining degree of entanglement can be quantified through the linear entropy $S_{\text{lin}}(\hat{\rho}_q) = 1 - \gamma_q$~\cite{agarwal_quantitative_2005,berrada_beam_2013}, where $\gamma_q = \Tr[\hat{\rho}_q^2]$ denotes the purity of the reduced density matrix $\hat{\rho}_q$ (see Supplementary Material)
\begin{equation}
	\gamma_q 
	= \int \dd \varepsilon_{\alpha}\! \int \dd \varepsilon_{\alpha}'
	\mathcal{Q}(\varepsilon_{\alpha})
	\mathcal{Q}(\varepsilon_{\alpha}')
	\abs{\braket{\varrho \, d_{\varepsilon_{\alpha}}(\omega_q)}{\varrho \, d_{\varepsilon_{\alpha}'}(\omega_q)}}^2.
\end{equation}
A value of $\gamma = 1$, or equivalently $S_{\text{lin}}(\hat{\rho}_q) = 0$, corresponds to a pure state $\hat{\rho}_q$, implying the absence of entanglement between the $q$-th harmonic and all other modes.~Under the assumption that the global quantum optical state remains pure, which is ensured in our treatment by the purity of the initial state [Eq.~\eqref{Eq:init:state}] and the unitary evolution [Eq.~\eqref{Eq:vNdiff}] (see Supplementary Material for an alternative derivation), the linear entropy provides a valid measure of entanglement between the $q$-th harmonic and the remaining modes.~If global purity were not guaranteed, reduced mixedness could also arise from correlations with an external system or from an initially mixed state, and the linear entropy would no longer constitute a faithful entanglement measure.

Figure~\ref{fig:correlations}~(e) shows the linear entropy $S_{\text{lin}}(\hat{\rho}_q)$ for different harmonic orders as a function of the two-color phase $\phi$.~Overall, we find a significant degree of entanglement that oscillates with $\phi$, reaching its maximum values for the even harmonics.~In contrast, for the odd orders, $S_{\text{lin}}(\hat{\rho}_q)$ remains close to zero, particularly within the plateau region $q_{\text{odd}}\in[9,21]$.~Beyond the HHG cutoff, around $q = 21$, the entanglement rapidly vanishes for both even and odd harmonics, as these modes approach a vacuum state.~The oscillatory behavior of $\gamma_q$ with $\phi$ originates from the phase-dependent interference between the $\omega$ and $2\omega$ driving fields, which modulates the instantaneous field intensity experienced by the medium~\cite{pedatzur2015attosecond}. In the presence of squeezing, these intensity fluctuations translate into phase-dependent quantum correlations between the driving field and the emitted harmonics. Higher harmonic orders, being more sensitive to intensity variations, therefore exhibit a stronger modulation of the linear entropy as a function of the relative phase.

Finally, the presence of entanglement between the $q$-th harmonic and the remaining modes is not incompatible with the absence of pair-wise harmonic entanglement.~Taken together, these observations indicate the emergence of multipartite correlations involving the driving field and multiple harmonic modes which, when reduced to single- or two-harmonic subsystems, manifest as reduced purity while remaining compatible with classically correlated mixtures at the pair-wise level.~In this context, our results demonstrate that driving HHG with bright quantum light enables controlled generation of correlated harmonics, with the $\omega-2\omega$ configuration providing a tunable mechanism to enhance or suppress these correlations via the two-color delay.~This conclusion, however, holds within the strong-field and low-depletion regimes considered here (see Supplementary Material for further discussion).~Beyond this regime, genuinely squeezed and entangled harmonic states may arise even when coherent states are used as drivers in gaseous~\cite{stammer2024entanglement}, semiconductor~\cite{theidel2024evidence}, and strongly correlated~\cite{lange2024electron} materials.

\subsection{Attosecond quantum tomography}\label{Sec:AQT}
\begin{figure}
	\centering
	\includegraphics[width=1\columnwidth]{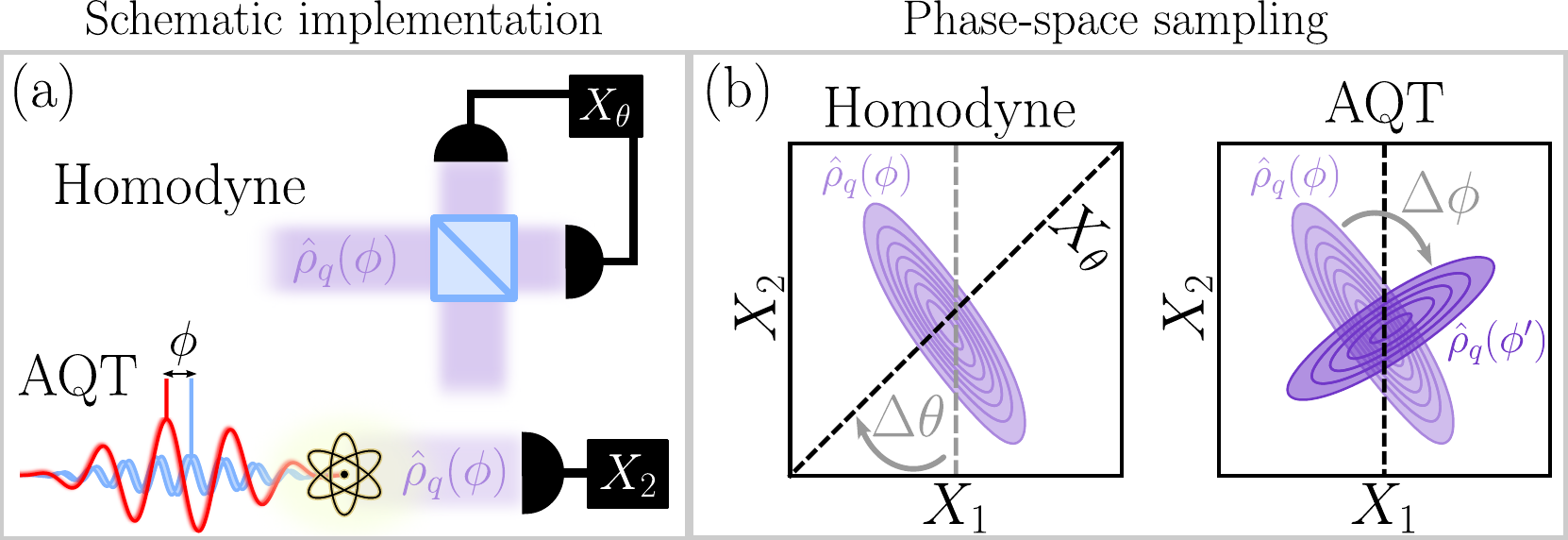}
	\caption{\textbf{Analogy between standard homodyne detection and Attosecond Quantum Tompgraphy}~(a) Schematic comparison between ideal homodyne tomography and AQT, highlighting their analogy as a phase-sensitive tomographic measurement.~(b) Schematic illustration of the corresponding phase-space sampling performed in each protocol. While homodyne detection measures a fixed state $\hat{\rho}_q(\phi)$ by rotating the quadrature orientation $X_\theta$, the AQT approach keeps the quadrature direction fixed and rotates the state $\hat{\rho}_q(\phi) \to \hat{\rho}_q(\phi')$. }
	\label{Fig:Homodyne:vs:AQT}
\end{figure}

Characterizing the quantum state of the harmonics in practice, however, remains extremely challenging.~The main difficulty arises from the lack of optical elements enabling quantum state tomography in the extreme-ultraviolet (XUV) regime.~Standard techniques such as homodyne detection require both a phase-stabilized coherent state source, serving as a local oscillator (LO) reference field, and linear optical elements operating efficiently at XUV wavelengths. In homodyne detection, the state of interest $\hat{\rho}_q$, is overlapped with the LO at a beam splitter~[Fig.~\ref{Fig:Homodyne:vs:AQT}~(a)], and from the measured intensities of the outgoing fields one reconstructs quadrature probability distributions, from which the Wigner function of $\hat{\rho}_q$ can be reconstructed~\cite{smithey_measurement_1993,lvovsky_continuous-variable_2009}~[Fig.~\ref{Fig:Homodyne:vs:AQT}~(b)].~Varying the phase of the LO corresponds to selecting different phase-space quadratures, thereby enabling tomographic reconstruction of the state.

Despite these experimental challenges, the $\omega-2\omega$ configuration considered here provides an alternative route for probing the quantum state of the high harmonics~[Fig.~\ref{Fig:Homodyne:vs:AQT}~(a)].~Motivated from the results shown in Fig.~\ref{fig:state_engineering}, the two-color delay has an analogous role to that of a LO: varying the delay $\phi$ rotates the harmonic quantum state $\hat{\rho}_q(\phi)$ in phase space, and measurements of the fixed quadrature operator $\hat{X} = \hat{a}+\hat{a}^\dagger$ allows to effectively probe the probability distribution $p(\phi) = \langle X | \hat{\rho}_q(\phi) | X \rangle$, with the quadrature eigenstates $|X \rangle$. This is inverse to standard homodyne detection, where the quantum state to be probed is fixed and the LO effectively changes the measured quadrature by changing its phase.~In contrast, in the present approach, the harmonic quantum state explicitly rotates with the two-color phase $\phi$, while the quadrature operator remains fixed in phase-space~[Fig.~\ref{Fig:Homodyne:vs:AQT}~(b)].~We understand this approach as \textit{attosecond quantum tomography (AQT)}.

\begin{figure}
	\centering
	\includegraphics[width=1\columnwidth]{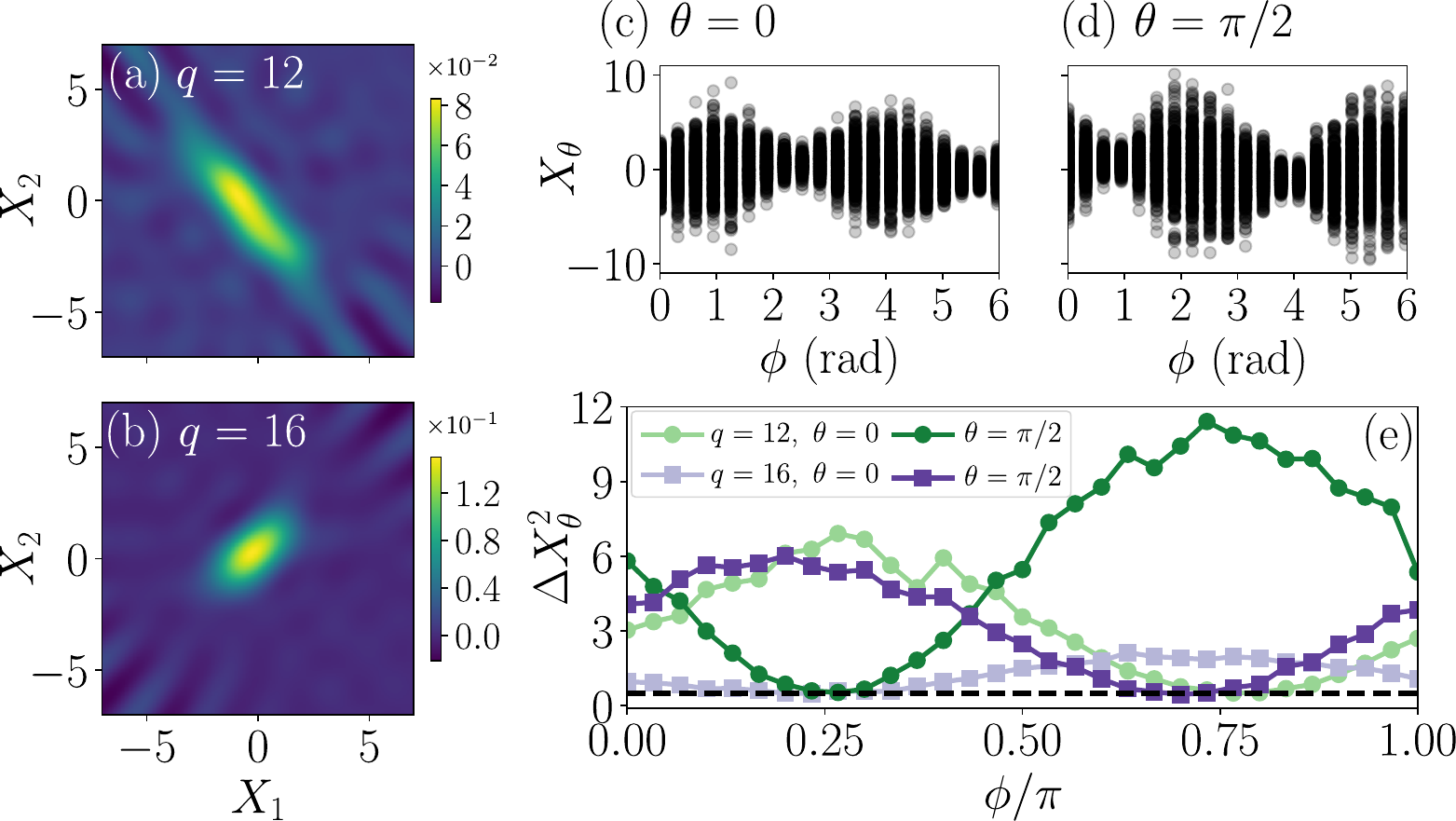}
	\caption{\textbf{Learning the quantum state through attosecond quantum tomography (AQT).}~(a), (b) Reconstructed AQT-distribution for the 12th and 16th harmonic orders with $\theta = 0$.~(c), (d)~AQT-traces for the 12th harmonic mode for two different quadrature operators $\hat{X}_{\theta} = \hat{a} e^{-i\theta} + \hat{a}^\dagger e^{i\theta}$.~(e) Variance of $\hat{X}_{\theta}$ for the 12th (green) and 16th (purple) harmonics, computed from the  AQT-traces for $\theta = 0$ (lighter curves) and $\theta = \pi/2$ (darker curves). The same parameters as in Fig.~\ref{fig:state_engineering} are used here.}
	\label{fig:tomography}
\end{figure}

To understand the AQT procedure, we perform a statistical sampling experiment numerically~\cite{rivera-dean_quantum_2025}: measurement outcomes $\{\lambda_i\}$ of $\hat{X}$ are sampled with probabilities $\{p(\lambda_i|\phi) = \bra{\lambda_i} \hat{\rho}_q(\phi)\ket{\lambda_i}\}$ of the state $\hat{\rho}_q(\phi)$ (see Supplementary Material). This sampling approach yields homodyne-like traces~[Fig.~\ref{fig:tomography}~(c),(d)], hereafter referred to as AQT-traces, from which a Wigner-like function of $\hat{\rho}_q$, the AQT-distribution, can be reconstructed [Fig.~\ref{fig:tomography}~(a),(b)] via an inverse Radon transformation~\cite{lvovsky_continuous-variable_2009}.~The resulting AQT-distribution exhibits quadrature-dependent variances, stretched along some directions and compressed along others, indicating squeezed like behavior, although the Heisenberg limit is never surpassed~[Fig.~\ref{fig:tomography}~(e)].

Nevertheless, we emphasize that the AQT approach does not constitute a genuine quantum state tomography method.~Unlike in standard homodyne detection, here, the effective LO is not independent of the probed state. The two-color delay influences both the measurement quadrature and modifies the photon statistics of the harmonics~[Fig.~\ref{fig:state_engineering}]. As a result, the reconstructed AQT-distributions are influenced by the same parameters that define the state [Fig.~\ref{fig:tomography}].~This dependence can be made more transparent by introducing the general quadrature operator $\hat{X}_{\theta} = \hat{a}e^{-i\theta} + \hat{a}^\dagger e^{i\theta}$ and noting that the evaluation of the AQT-traces for different $\theta$ values yields distinct results.~Figure~\ref{fig:tomography} illustrates this for the 12th harmonic order: beyond a trivial phase delay, the field fluctuations are noticeably larger for $\theta = \pi/2$ [Fig.~\ref{fig:tomography}~(b)] than for $\theta = 0$ [Fig.~\ref{fig:tomography}~(a)]. Consequently, the harmonic properties vary with $\theta$, as shown in Fig.~\ref{fig:tomography}~(e) for two harmonic orders: a change of the fixed reference quadrature results in different distributions, i.e., the reconstructed state depends on the measurement basis.

This highlights the key limitation of this approach. The properties of the effective LO strongly influence the reconstructed harmonic states, while in any true tomography method the probe should not change the signal itself.~Nonetheless, given the practical challenges associated with implementing standard quantum optical tomography techniques in the XUV regime, it is instructive to evaluate the extent to which AQT accurately reproduces the results obtained via conventional homodyne detection.~To this end, we perform this comparison in two complementary ways: first, by comparing the AQT-distribution and the proper homodyne-based Wigner function; and second, by comparing the corresponding quadrature probability distributions.

\begin{figure}
	\centering
	\includegraphics[width=1\columnwidth]{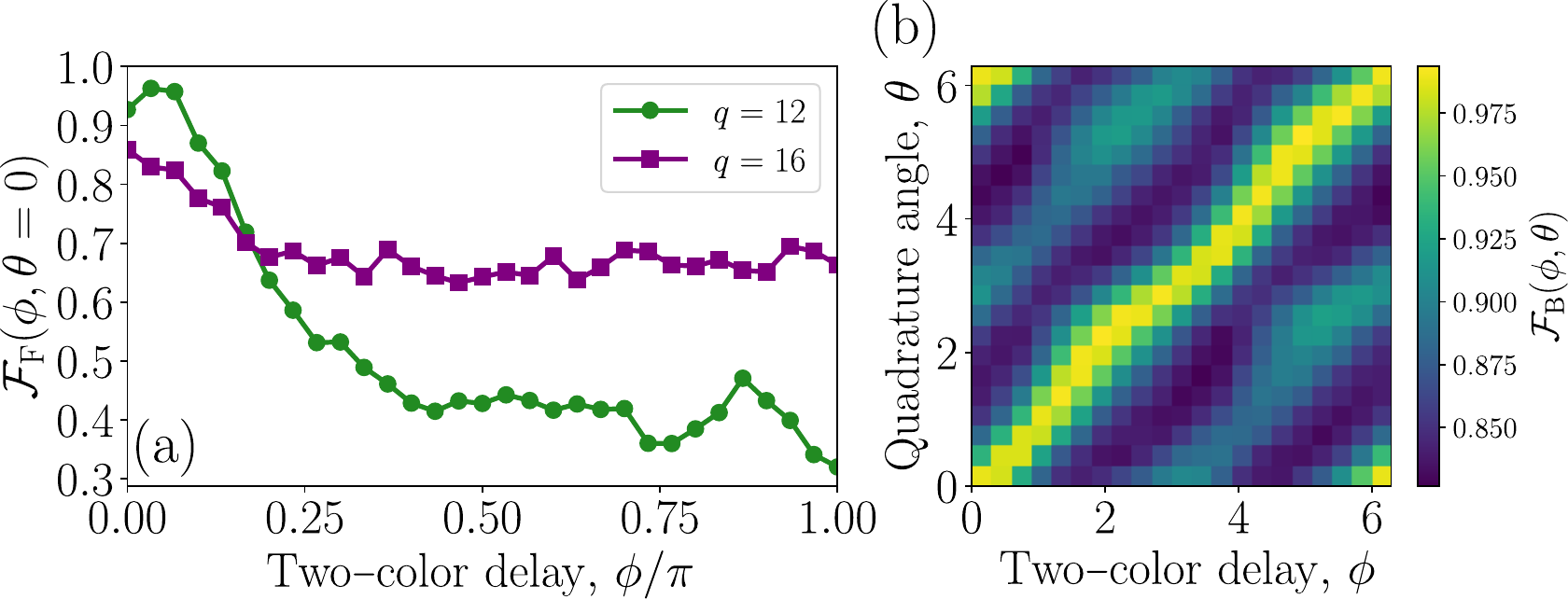}
	\caption{\textbf{Fidelity of AQT with respect to homodyne detection.}~(a) Frobenius norm between the Wigner functions reconstructed via AQT (with $\theta=0$) and exact homodyne tomography, shown as a function of the two-color delay $\phi$ for two different harmonics $q\in \{12,16 \}$.~(b) Bhattacharyya distance between the exact homodyne traces and the AQT-traces obtain for the 12th harmonic, evaluated while varying both the quadrature angle $\theta$ and the two-color delay $\phi$.~The same parameters as those used in Fig.~\ref{fig:state_engineering} are used here.}
	\label{Fig:Fidelities}
\end{figure}

For the first comparison, we note that Wigner function reconstructions yield finite-dimensional matrices $\mathsf{W}$, whose size depends on the chosen phase-space discretization.~We therefore compare the AQT-distributions $\mathsf{W}^{(\text{AQT})}(\theta)$ with the homodyne-reconstructed true Wigner function $\mathsf{W}^{(\text{H})}(\phi)$, the latter yielding Fig.~\ref{fig:state_engineering}~(a)-(h) under ideal conditions.~Here, we emphasize that $\mathsf{W}^{(\text{AQT})}(\theta)$ depends on the optical quadrature $\hat{X}_{\theta}$ selected in the AQT protocol through its dependence with $\theta$~[Fig.~\ref{fig:tomography}~(c),(d)], whereas $\mathsf{W}^{(\text{H})}(\phi)$ depends on the quantum state $\hat{\rho}(\phi)$ and therefore varies with the two-color phase $\phi$. To quantify the agreement between both reconstructions, we use the normalized Frobenius inner product as a fidelity measure
\begin{equation}
	\mathcal{F}_{\text{F}}(\phi,\theta)
		= \dfrac{\sum_{i,j} \big[ \mathsf{W}^{(\text{AQT})}_{i,j}(\theta)\mathsf{W}^{(\text{H})}_{i,j}(\phi)\big]}{\sqrt{\sum_{i,j} \big[\mathsf{W}^{(\text{AQT})}_{i,j}(\theta)\big]^2}\sqrt{\sum_{i,j} \big[\mathsf{W}^{(\text{H})}_{i,j}(\phi)\big]^2}}.
\end{equation}

For the second comparison, we consider the quadrature probability distributions $\{P_{\text{AQT}}(X|\phi;\theta)\}$ and $\{P_{\text{H}}(X|\theta;\phi)\}$, obtained from the AQT and homodyne traces, respectively, using a Gaussian kernel density estimator~\cite{silverman_density_2018}. In this case, either the quadrature angle $\theta$ (for AQT) or the two-color phase $\phi$ (for homodyne) is held fixed while the other parameter is varied.~The similarity between both sets of distributions is quantified using the Bhattacharyya coefficient~\cite{bhattacharyya1946measure}
\begin{equation}
	\mathcal{F}_{\text{B}}(\phi,\theta)
		= \dfrac{1}{N}
				\sum^{N}_{i=1}\int \!\dd X
					 \sqrt{P_{\text{AQT}}(X|\phi_i;\theta) 
					 	P_{\text{H}}(X|\theta_i;\phi)},
\end{equation} 
where $N$ denotes the number of angles sampled along the horizontal axis of Figs.~\ref{fig:tomography}~(c),(d).~Figure~\ref{Fig:Fidelities} displays the results of this analysis.~For both fidelity measures, values close to unity are obtained whenever $\theta \approx \phi$, corresponding to maximal agreement between the AQT and homodyne reconstructions.~This correspondence is evident when comparing Fig.~\ref{fig:scheme}~(a),(b) with Fig.~\ref{fig:tomography}~(a),(b), where the phase-space distributions overlap most closely. 

For the Frobenius-based fidelity, the agreement decreases as the difference $\abs{\phi-\theta}$ (defined modulo $2\pi$) increases.~Varying the two-color phase $\phi$ modifies not only the degree of phase-space stretching of the Wigner function, but also the position of its center, which oscillates around the origin as $\phi$ varies from $0$ to $2\pi$.~These combined effects reduced the overlap between the AQT distribution and the homodyne-reconstructed Wigner function, thereby explaining the differences between the two considered harmonic orders: the 12th harmonic exhibits more pronounced stretching than $q=16$ (see Fig.~\ref{fig:state_engineering}), further diminishing the overlap between $\mathsf{W}^{(\text{AQT})}(\theta)$ and $\mathsf{W}^{(\text{H})}(\phi)$.~Consequently, $	\mathcal{F}_{\text{F}}(\phi,\theta)$ reaches a minimum near $\phi =\pi$, where both distributions are stretched along the same-phase space direction, but are displaced relative to one another (see Fig.~\ref{fig:tomography}~(a),(b) and Fig.~\ref{fig:scheme}~(d),(h)). In contrast, the Bhattacharyya-based fidelity yields overall higher values.~Although it decreases when $\theta \neq \phi$, a secondary local maximum appears near $\theta = \phi + \pi$.~Under these conditions, the mean values of the AQT and homodyne traces are shifted in phase space, yet their fluctuation profiles remain aligned.~As a result, the overlap between the probability distributions $P_{\text{AQT}}(X,\phi;\theta)$ and $P_{\text{H}}(X,\theta;\phi)$ partially revives, leading to an increased  $\mathcal{F}_{\text{B}}(\phi,\theta)$.

Despite the differences, the overall analysis demonstrates that AQT can be carefully engineered to reconstruct the harmonic quantum state $\hat{\rho}_q(\phi)$ with high fidelity, typically at or above the 90\% fidelity level.~Given the formidable challenge of generating independent, phase-stable LOs in the XUV regime, the introduced AQT approach provides a crucial bridge between theory and experiment. This offers practical means to benchmark quantum-optical models of HHG and explore the limits of strong-field quantum optics.~Equally important, for classical driving fields the AQT procedure provides insight into how the $2\omega$ component influences the electron quantum trajectories, which can be inferred from the intensity ratio between even and odd harmonic orders~\cite{pedatzur2015attosecond}.~In contrast, this method ceases to be effective when the driving field carries quantum fluctuations, such as squeezing.~In that case, homodyne-based techniques offer a viable alternative for probing the underlying electron dynamics (see Supplementary Material).

\section{\label{sec:conclusion}DISCUSSION}
We conclude by summarizing our main findings, situating them within recent experimental developments in the nascent field of \emph{Extreme Quantum Optics}, and outlining future perspectives.

\subsection{Summary}
This work has explored the concept of \textit{attosecond quantum interferometry}, developing its capabilities to control the phase-space properties, photon statistics and correlations of the HHG-emitted radiation, and indicates its inherent limitations as a quantum state tomography technique, while simultaneously allow for new insights onto learning the quantum state.~While the two-color phase delay plays a role analogous to that of a local oscillator in homodyne detection, it simultaneously alters the quantum properties of the state itself.~Consequently, the outcomes of the homodyne-like measurements obtained through AQT do not yield proper reconstructions of the harmonic quantum state.~Nonetheless, they provide robust and quantitatively reliable access to its phase-space properties, yielding reconstruction fidelities approaching 90\% or higher compared to those obtained through homodyne detection.

Thus, given the practical difficulty of performing true quantum state tomography in the extreme-ultraviolet regime, AQT  emerges as a valuable alternative for probing the phase-space properties of the emitted radiation. Furthermore, our analysis shows that AQT can act as a witness of non-classical fluctuations in the driving field. For instance, when the $2\omega$ component is thermal, the variance $(\Delta X_{\theta})^2 \gg 0.5$ for all quadrature angles $\theta$, unlike to what is found for the case of squeezed driving fields where $\min_{\theta}[(\Delta X_{\theta})^2] \simeq 0.5$.~Thus, the sensitivity of the harmonic radiation properties to the nature of the driving field make them an ideal probe to characterize the properties of the drive.

\subsection{Relation to recent experimental advances}
Recent experimental advances at the intersection of quantum optics and ultrafast strong-field physics have begun to address the challenge of characterizing the properties of the emitted radiation via HHG. Of particular relevance to the theoretical framework developed here are the experimental results reported in Ref.~\cite{tzur2025measuring}, where a closely related configuration was investigated, albeit with the roles of the $\omega$ and $2\omega$ fields interchanged relative to our scheme, resulting in a different set of selection rules for the emitted harmonics.~In that study, it was demonstrated that squeezed fluctuations of the driving field can be transferred to the ultrafast electron dynamics underlying HHG, building on concepts originally developed for classical two-color drivers~\cite{pedatzur2015attosecond}.~On this basis, the authors implemented a tomographic-like reconstruction of the emitted harmonics, based on the AQT framework introduced here.

Our theoretical framework complements and generalizes these findings by providing a quantitative analysis of the phase-space structure, reconstruction fidelity, and intrinsic limitations, thus clarifying the operational boundaries of such tomographic-like schemes as genuine quantum state characterization protocols.~Taken together, these theoretical and experimental developments indicate that the AQT approach developed here is emerging as a practically viable and conceptually powerful tool for probing the phase-space properties of the emitted HHG radiation in regimes where conventional quantum optics techniques remain experimentally challenging.

Finally, related developments within nonlinear optics could further illustrate the potential relevance of these ideas towards ultrafast physics. Recent work has demonstrated quantum state characterization of squeezed light in a nonlinear degenerate four-wave mixing (FWM) process, when driven by two spectrally distinct classical ultrafast pulses with partial spectral overlap in the interaction region~\cite{sennary_attosecond_2025}.~In that case, the properties of the emitted radiation could be indirectly reconstructed through comparison with those of the driving fields.~Although such a mapping is more challenging in HHG due to the large spectral separation between the driver and the emitted radiation, it remains an intriguing question whether controlled phase relations between the two coherent pump fields could be map on to the phase-space properties of the FWM-generated light, in a manner analogous to AQI. Exploring this possibility could extend AQI-like methodologies beyond HHG, towards ultrafast nonlinear optics.

\subsection{Outlook}
The approach for engineering the photon statistics and phase-space properties of the emitted light developed in this work, opens up exciting prospects for future studies, particularly in the context of entangled driving fields~\cite{iskhakov_polarization-entangled_2012}.~In such cases, the quantum correlations in the driving field restricts the applicability of the classical limits, which would otherwise obscure their intrinsic non-classical features.~Our findings therefore motivate the exploration of whether and how quantum correlations in the driving field can be transferred to the emitted harmonics, potentially enabling the generation of multimode entangled states of light in the XUV regime. 

A central question in this context is whether AQI and AQT can faithfully preserve and quantify genuine non-classical properties of the emitted radiation under conditions in which such features cannot be neglected, for instance beyond the low-depletion regime~\cite{stammer2024entanglement}, or in the presence of intrinsic material correlations within the generating medium~\cite{theidel2024evidence,lange2024electron}.~In particular, given that violations of the CSI have been identified even for classically driven configurations~\cite{theidel2024evidence}, it remains to be understood whether additional correlations introduced by strongly squeezed or entangled driving fields would enhance such violations or instead redistribute and possibly dilute them.

Looking ahead to practical implementations, ultrafast physics has long demonstrated attosecond precision in the spectroscopy of a wide range of materials~\cite{marangos_development_2016,ghimire_high-harmonic_2019,zong_emerging_2023}. In parallel, the use of quantum light has emerged as a powerful route toward enhanced resolution and selective manipulation in spectroscopic protocols~\cite{schlawin_pump-probe_2016,dorfman_nonlinear_2016,mukamel_roadmap_2020}, offering control parameters that are fundamentally complementary to those available in attosecond strong-field techniques. Given the intrinsically pump-probe nature of many ultrafast spectroscopies, AQI may provide a natural interface between these two domains, enabling the development of extreme quantum optics spectroscopy schemes in which attosecond temporal resolution is combined with genuinely non-classical light sources.

\begin{acknowledgments}

P.S. acknowledges funding from: The European Union’s Horizon 2020 research and innovation programme under the Marie Skłodowska-Curie grant agreement No 847517.
ICFO-QOT group acknowledges support from:
European Research Council AdG NOQIA; MCIN/AEI (PGC2018-0910.13039/501100011033,  CEX2019-000910-S/10.13039/501100011033, Plan National STAMEENA PID2022-139099NB and FUNQIP PID2022-139658NB-I00, project funded MCIN and  by the “European Union NextGenerationEU/PRTR" (PRTR-C17.I1), FPI); QUANTERA DYNAMITE PCI2022-132919, QuantERA II Programme co-funded by European Union’s Horizon 2020 program under Grant Agreement No 101017733; Ministry for Digital Transformation and of Civil Service of the Spanish Government through the QUANTUM ENIA project call - Quantum Spain project, and by the European Union through the Recovery, Transformation and Resilience Plan - NextGenerationEU within the framework of the Digital Spain 2026 Agenda; MICIU/AEI/10.13039/501100011033 and EU (PCI2025-163167); Fundació Cellex;
Fundació Mir-Puig; Generalitat de Catalunya (European Social Fund FEDER and CERCA program; Barcelona Supercomputing Center MareNostrum (FI-2023-3-0024); 
Funded by the European Union (HORIZON-CL4-2022-QUANTUM-02-SGA, PASQuanS2.1, 101113690, EU Horizon 2020 FET-OPEN OPTOlogic, Grant No 899794, QU-ATTO, 101168628),  EU Horizon Europe Program (No 101080086 NeQSTGrant Agreement 101080086 — NeQST). J.~R.-D.~acknowledges funding from  UKRI2300 - Attosecond Photoelectron Imaging with Quantum Light.

\smallskip

\end{acknowledgments}

\bibliography{literatur}{}

\appendix
\clearpage
\onecolumngrid

\begin{center}
    {\large \textbf{\textsc{Supplementary Material}}}
\end{center}

\section{About the change of variables}\label{Sec:App:Change:Variables}
Here, we consider a scenario in which the squeezing parameter $\xi(\phi)$ itself varies with $\phi$, opposite to previous work where this was fixed~\cite{stammer_weak_2025}. In this way, the $2\omega$ field consistently exhibits the same type of squeezing, regardless of the phase difference $\phi$. We can therefore express the initial quantum optical state of the $\omega$ and $2\omega$ drivers as
\begin{equation}
	\ket{\Psi(t_0)}
	= \ket{\alpha_{\omega}}
	\otimes \hat{D}(\alpha_{2\omega}) \hat{S}(\xi) \ket{0},
\end{equation}
where we have $\alpha_{2\omega} = \abs{\alpha_{2\omega}} e^{i\phi}$ for the second harmonic amplitude, with $\abs{\alpha_{2\omega}} = \epsilon \abs{\alpha_{\omega}}$ and $\epsilon \ll 1$. To determine the relation $\xi(\phi)$ as well as providing a more convenient coordinate system for our calculations, we evaluate the Husimi function of the $2\omega$ field, since this quantity will be required for our subsequent calculations. Writing $\xi = r e^{i\bar{\theta}}$, with $r > 0$, the Husimi function can be expressed as
\begin{equation}\label{Eq:Husimi}
	\begin{aligned}
		Q(\alpha)
		&= \dfrac{1}{\pi\cosh r}
		\exp[-\abs{\bar{\alpha}}^2
		- \dfrac{\tanh{r}}{2}
		\big(
		e^{i\bar{\theta}} \bar{\alpha}^{*2}
		+ e^{-i\bar{\theta}} \bar{\alpha}^2
		\big)]
		\\&
		= \dfrac{1}{\pi\cosh r}
		\exp[-\dfrac{\gamma^2_xe^r}{\cosh(r)}
		-\dfrac{\gamma^2_ye^{-r}}{\cosh(r)}],
	\end{aligned}
\end{equation}
where we define $\bar{\alpha} = \alpha - \alpha_{2\omega}$ and $\gamma = \bar{\alpha}e^{-i\bar{\theta}/2}$. 
Interestingly, in the frame of reference defined by $\gamma$, the squeezing parameter is locally invariant; that is, the type of squeezing remains unchanged.~As a result, taking the classical limit becomes straightforward. The relation between the frame of reference defined by $\alpha$ and that defined by $\gamma$ is 
\begin{equation}
	\alpha = \gamma e^{i\bar{\theta}/2} + \abs{\alpha_{2\omega}} e^{i\phi},
\end{equation}
which, when written in terms of their real and imaginary parts, becomes
\begin{equation}
	\begin{aligned}
		&\alpha_x 
		= \gamma_x \cos(\bar{\theta}/2)
		+ \abs{\alpha_{2\omega}} \cos(\phi)
		- \gamma_y \sin(\bar{\theta}/2),
		\\&
		\alpha_y
		= \gamma_x \sin(\bar{\theta}/2)
		+ \abs{\alpha_{2\omega}} \sin(\phi)
		+ \gamma_y \cos(\bar{\theta}/2).
	\end{aligned}
\end{equation}

\begin{figure}[h!]
	\centering
	\includegraphics[width=1\textwidth]{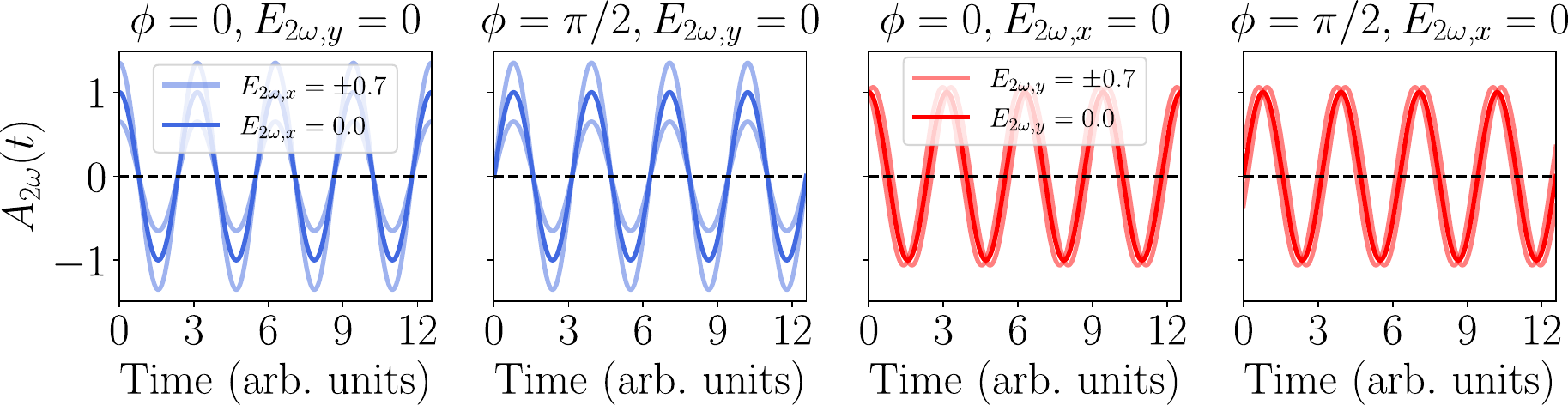}
	\caption{Vector potential dependence with time in the original frame of reference. Here, we change the values of $E_{2\omega,x}$ and $E_{2\omega,y}$, defined in through $\gamma_x$ and $\gamma_y$, while keeping the mean field strength constant.}
	\label{Fig:fields:example}
\end{figure}

To guarantee that the type of squeezing remains constant as $\phi$ varies, we set $\bar{\theta} = 2 \phi$. In this case, the relation between the old and new integration variables is
\begin{equation}\label{Eq:change:variables}
	\begin{aligned}
		&\alpha_x 
		= (\gamma_x + \abs{\bar{\alpha}})\cos\phi
		- \gamma_y \sin\phi,
		\\
		&\alpha_y
		= (\gamma_x + \abs{\bar{\alpha}})\sin\phi
		+ \gamma_y \cos\phi,	
	\end{aligned}
\end{equation}
whose effect is illustrated in Fig.~\ref{Fig:fields:example}. Specifically, we show the vector potential of the $2\omega$ field, given by
\begin{equation}
	A_{2\omega}(t)
	= \dfrac{\bar{E}_{2\omega,x}}{2\omega} \cos(2\omega t)
	+ \dfrac{\bar{E}_{2\omega,y}}{2\omega} \sin(2\omega t),
\end{equation}
where $\bar{E}_{2\omega,y}$ denotes the electric field strength in the original frame of reference, i.e., the one defined by $\alpha$. In contrast, in th $\gamma$-frame we vary $E_{2\omega,x}$ (in blue) and $E_{2\omega,y}$ (in red), while keeping the mean field strength fixed, thereby producing the desired phase and amplitude squeezing effects.

\smallskip
A natural, and perhaps trivial, question at this stage is whether the change of variables affects how the classical limit is taken.~This limit acts, effectively, on the $P(\alpha,\beta^*)$ function when evaluating physical observables of the harmonics~\cite{stammer_weak_2025}.~Consequently, the analysis presented here concerns solely the $P(\alpha,\beta^*)$ function, which can be generally written as~\cite{d_drummond_quantum_2016}
\begin{equation}
	P(\alpha,\beta^*)
	= \dfrac{1}{4\pi}
	\exp[-\dfrac{\abs{\alpha-\beta^*}^2}{2}]
	Q
	\bigg(
	\dfrac{\alpha+\beta^*}{2}
	\bigg),
\end{equation}
and we are interested in evaluating integrals of the form
\begin{equation}\label{Eq:integral}
	I = \int \dd^2 \alpha \int \dd^2 \beta P(\alpha,\beta^*)o(\alpha,\beta^*),
\end{equation}
where $o(\alpha,\beta^*)$ is a well-behaved function whose specific form depends on the observable under consideration. Since we generally write the Husimi function as in Eq.~\eqref{Eq:Husimi}, the change of variables relevant to our case reads
\begin{equation}\label{Eq:change:variables:2}
	\begin{aligned}
		&\dfrac{\alpha_x +\beta_x}{2}
		= (\gamma_x + \abs{\bar{\alpha}})\cos\phi
		- \gamma_y \sin\phi,
		\\
		&\dfrac{\alpha_y -\beta_y}{2}
		= (\gamma_x + \abs{\bar{\alpha}})\sin\phi
		+ \gamma_y \cos\phi.	
	\end{aligned}
\end{equation}

Importantly, for single-mode squeezed states, the $P(\alpha,\beta^*)$ function can be written as
\begin{equation}
	P(\alpha,\beta*)
	= \dfrac{1}{4\pi^2\cosh(r)}
	f_1\bigg(\dfrac{\alpha_x-\beta_x}{2}\bigg)
	f_2\bigg(\dfrac{\alpha_y+\beta_y}{2}\bigg)
	g\bigg(\dfrac{\alpha_x+\beta_x}{2},
            \dfrac{\alpha_y-\beta_y}{2}
        \bigg),
\end{equation}
namely, as a product of three functions, each of them depending on variables of a linearly independent set. Under the transformation
\begin{equation}
	x_+ = \dfrac{\alpha_x + \beta_x}{2},\
	x_{-} = \dfrac{\alpha_x - \beta_x}{2},\
	y_+ = \dfrac{\alpha_y + \beta_y}{2},\
	y_{-} = \dfrac{\alpha_y - \beta_y}{2},
\end{equation}
we can therefore rewrite Eq.~\eqref{Eq:integral} as
\begin{equation}
	I = \dfrac{1}{\pi^2 \cosh(r)}
	\int \dd x_+ \int \dd x_-
	\int \dd y_+ \int \dd y_-
	f_1(x_-)f_2(y_+)g(x_+,y_-)
	o(x_+,x_-,y_+,y_-),
\end{equation}
with the integration limits unaffected, since they span the entire $\mathbbm{R}^4$.~In terms of these new variables, we may further introduce the more convenient change of variables defined in Eq.~\eqref{Eq:change:variables:2}, so that the integral reads
\begin{equation}
	I = \dfrac{1}{\pi^2 \cosh(r)}
	\int \dd x_- \int \dd \gamma_x \int \dd \gamma_y \int \dd y_+
	\exp[-x_-^2]\exp[-y_+^2]
	\exp[-\dfrac{\gamma_x^2e^r}{\cosh(r)}]
	\exp[-\dfrac{\gamma_y^2e^{-r}}{\cosh(r)}]
	g(x_-,\gamma_x,\gamma_y,y_+),
\end{equation}
and allows factorizing $g(x_+,y_-) = g_1(x_+)g_2(y_-)$. Here, the classical limit can be applied in the usual manner.

\section{Evaluation of quantum optical observables}\label{Sec:App:QO:obs}

In the low-depletion regime, the final quantum optical state associated to the $q$th harmonic mode, after the interaction with an ensemble of $N$ atoms~\cite{stammer2023quantum}, can be expressed as
\begin{equation}\label{Eq:Meth:state:2}
	\begin{aligned}
		\hat{\rho}_q(t)
		= \int \dd^2\alpha \int \dd^2 \beta& \dfrac{P(\alpha,\beta^*)}{\braket{N\chi_{\beta^*,q}(t)}{N\chi_{\alpha,q}(t)}}
		\dyad{N\chi_{\alpha,q}}{N\chi_{\beta^*,q}},
	\end{aligned}
\end{equation}
from which any quantum optical observable $\hat{O}$ acting on the harmonic mode $q$ can be computed as
\begin{equation}\label{Eq:Meth:Obs:2}
	\begin{aligned}
		\langle\hat{O}\rangle_q
		= \int \dd^2\alpha \int \dd^2 \beta&
		\dfrac{P(\alpha,\beta^*)}{\braket{N\chi_{\beta^*,q}(t)}{N\chi_{\alpha,q}(t)}}
		o(N\chi_{\alpha,q},N\chi_{\beta^*,q}),
	\end{aligned}
\end{equation} 
where we assume that the observable $\hat{O}$ does not introduce any additional dependencies on either the number of emitters $N$ nor the light-matter coupling parameter $\kappa$.

To evaluate these quantum optical observables, we work both in the classical and quasi-thermodynamic limits, defined as follows:
\begin{itemize}
	\item \textbf{Classical limit.}~In this regime, we express the coherent state amplitude $\varepsilon_{\alpha} = 2\kappa \alpha$, where $\varepsilon_\alpha$ denotes the electric field amplitude. This limit entails setting $V\to \infty$ and $\kappa \to \infty$, the first motivated by the fact that we are dealing with fields propagating in free space, where the quantization volume $V\to\infty$ (implying $\kappa\to0$). Consequently, to maintain a finite electric field amplitude, one must take $\alpha \to \infty$.
	\item \textbf{Quasi-thermodynamic limit.}~Since $V\to \infty$, a non-vanishing harmonic generation signal requires $N\to \infty$, such that $\varrho = \kappa N $ remains finite. As a result, the coherent state amplitude associated with the harmonic mode is given by $\chi_q = \sqrt{q} N\kappa \langle d(\omega_q)\rangle = \sqrt{q}\varrho\langle d(\omega_q)\rangle\equiv \varrho_q\langle d(\omega_q)\rangle$. It is important to note that this limit makes the coherent state amplitude $\varrho\langle d(\omega_q)\rangle$ constant, while the local intensity of the generated harmonics tends to zero. However, integrating the intensity over the entire generating volume gives a final total intensity $\int \dd \vb{r} I(\vb{r})  = \varrho^2\abs{\langle d(\omega_q)\rangle}^2 = \text{constant}$. 
\end{itemize}

Under these conditions, Eq.~\eqref{Eq:Meth:Obs:2} becomes
\begin{equation}\label{Eq:App:Obs:lim}
	\begin{aligned}
		\langle \hat{O}_q\rangle
		&= 
		\int \dd^2 \varepsilon_\alpha
		\int \dd^2 \varepsilon_\beta
		\bigg[
		\lim_{\kappa\to0} \dfrac{1}{16\kappa^4}P(\varepsilon_\alpha,\varepsilon^*_\beta)
		\bigg]
		\dfrac{o(\varrho_q\langle d_\alpha(\omega_q)\rangle,\varrho_q\langle d_{\beta^*}(\omega_q)\rangle)}{\braket{\varrho_q\langle d_{\beta^*}(\omega_q)\rangle}{\varrho_q\langle d_\alpha(\omega_q)\rangle}},
	\end{aligned}
\end{equation}
where, in the case of using squeezed light, the limiting behavior of the distribution $P(\alpha,\beta^*)$ is given by~\cite{even_tzur_photon-statistics_2023,rivera2025structured}
\begin{equation}
	\begin{aligned}
		\mathcal{Q}(\varepsilon_{\alpha})
		=
		\lim_{\kappa\to0} \dfrac{1}{16\kappa^4}P(\varepsilon_\alpha,\varepsilon^*_\beta)
		&= \dfrac{1}{\sqrt{2\pi \varsigma_i}}
		\exp[-\dfrac{\big(\varepsilon_{\alpha,i}-\bar{\varepsilon}_{i}\big)}{2\varsigma_i}]
		\delta(\varepsilon_\alpha - \varepsilon_\beta^*)
		\delta(\varepsilon_{\alpha,\bar{i}} - \bar{\varepsilon}_{\bar{i}}).
	\end{aligned}
\end{equation}
Here, $\bar{i}$ denotes the phase-space direction along which the squeezing is applied, $i$ the orthogonal direction, $\bar{\varepsilon}_{i}$ and $\bar{\varepsilon}_{\bar{i}}$ the coherent state amplitudes along each respective axis, and $\varsigma_i = 4 I_{\text{squ}}$ quantifies the increased field fluctuations along direction $i$.

\begin{figure}
    \centering
    \includegraphics[width=0.9\textwidth]{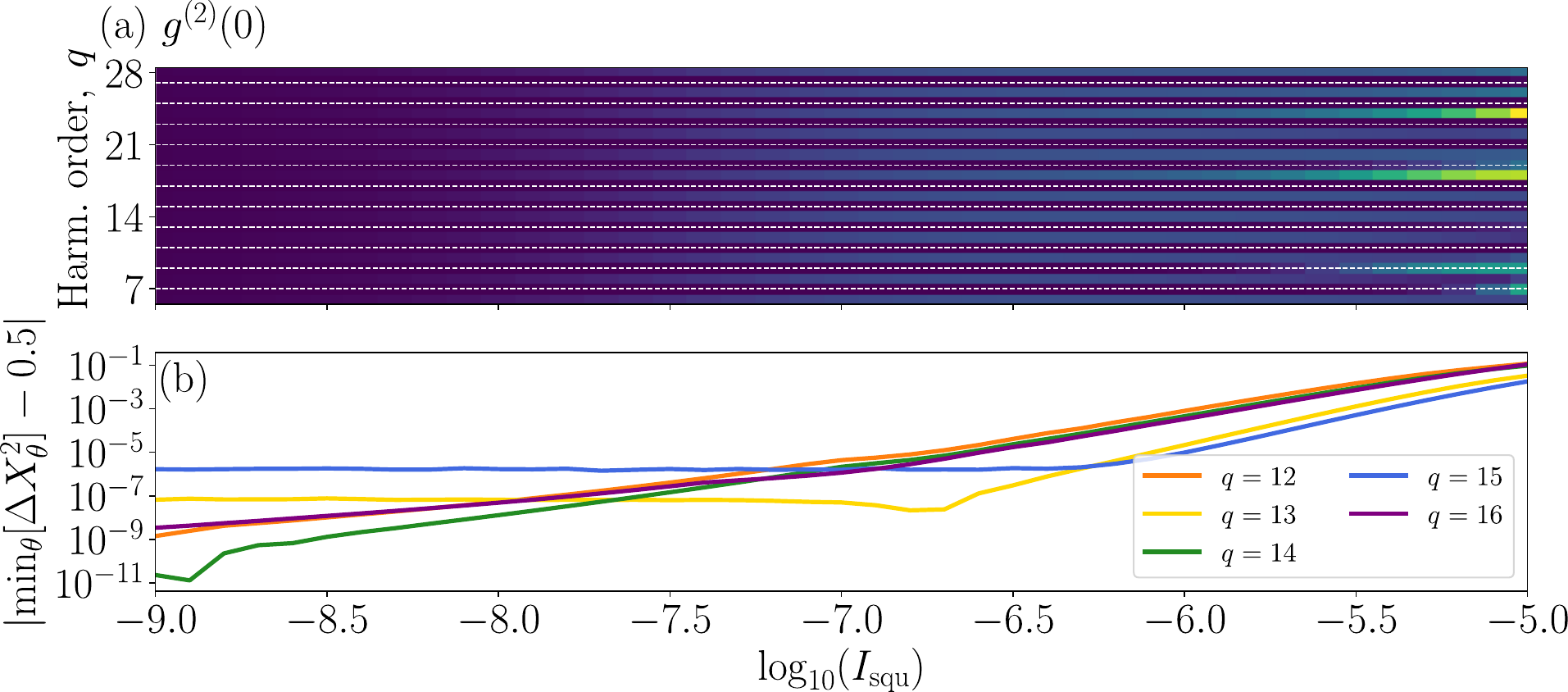}
    \caption{(a) Second-order autocorrelation function for different harmonic orders as a function of the squeezing intensity, with $\phi = 0$. Here, the white dashed lines indicate the odd harmonic orders.~(b) Difference between the variance of the optical quadrature along the squeezed direction and that of a coherent state.}
    \label{fig:g2:var:vs:Isqu}
\end{figure}

Figure~\ref{fig:g2:var:vs:Isqu} displays how the amount of squeezing present in the driver significantly affecrs the priperties of the emitted radiation. Interestingly, the minimum value of the variances remains quite close to the lower bound of 0.5 for a wide range of $I_{\text{squ}}$, suggesting the potential---highlighted in the main text---of using the properties of the harmonics to characterize those of the driving field. We note that these modifications arise because squeezing intensity enhances the noise fluctuations of the generated harmonics, as well, although also their mean response. In this context, we expect that when changing the atomic system under consideration---which alters the medium-dependent harmonic response---would primarily induce a displacement of the Wigner functions obtained in Fig.~2 of the main text. However, it should not significantly modify the optical quadrature variances or the generated photon statistics, which are predominantly governed by $I_{\text{squ}}.$

\begin{figure}[h!]
    \centering
    \includegraphics[width=1\textwidth]{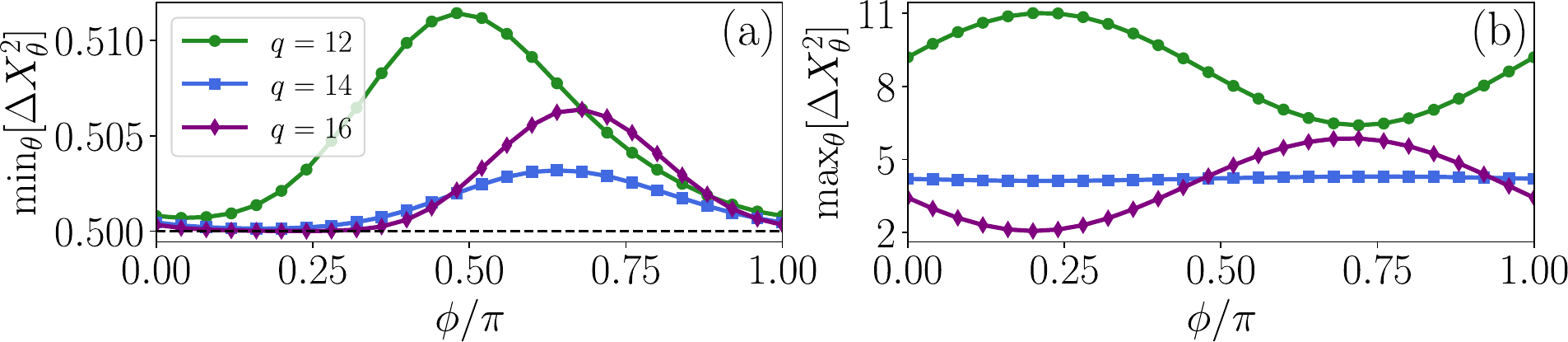}
    \caption{(a) Minimum and maximum variance fluctuations obtained when using phase-squeezed states instead of the amplitude-squeezed states considered in the main text. All other field parameters are kept the same as those leading to the results in Fig.~2 of the main text.}
    \label{Fig:Variances:Ph:Squ}
\end{figure}

Finally, while we focus in this manuscript on the use of amplitude squeezing, our results are also valid for the case where phase-squeezing is applied instead. In Fig.~\ref{Fig:Variances:Ph:Squ} we show results analogous to those presented in Fig.~2~(i) and (j) of the main text, when considering phase-squeezing. As observed, the features obtained are analogous to those found with amplitude-squeezed states, although shifted in phase by $\pi/2$. This is expected, due to the intrinsic $\pi/2$ phase difference on the field fluctuations leading to amplitude and phase squeezing.

\section{About the absence of genuinely squeezed states in the harmonic radiation}
It is of interest, however, to analyze whether the absence of squeezing in the generated harmonic modes is an intrinsic limitation of the theoretical analysis under consideration or a physical consequence of the strong-field regime in which we operate. To address this question explicitly, we consider a slightly modified theoretical analysis in which the initial driving field is expanded in the coherent state basis, namely
\begin{equation}
	\ket{\Psi(t_0)}
		= \dfrac{1}{\pi} \int \dd^2 \alpha_{2\omega} \braket{\alpha_{2\omega}}{\Phi} \ket{\text{g}}\otimes \ket{\alpha_\omega} \otimes \ket{\alpha_{2\omega}} \bigotimes_{q\geq 3} \ket{0_{q\omega}}, 
\end{equation}
where $\ket{\Phi} = \equiv \hat{D}(\alpha_{2\omega})\hat{S}(\chi)\ket{0_{2\omega}}$.~We evolve this state according to the dynamics generated by the Hamiltonian $\hat{H}(t) = \hat{H}_{\text{at}} + \mathsf{e} \hat{r}\cdot \hat{E}(t)$, where $\hat{H}_{\text{at}}$ denotes the atomic Hamiltonian and $\hat{E}(t) = -i\sum_q \kappa_q [\hat{a}_q e^{-i\omega_q t} + \text{h.c.}]$.~For simplicity, we restrict the discussion to the case of linear polarization. At any time $t \geq t_0$, the state can be written as
\begin{equation}
	\begin{aligned}
	\ket{\Psi(t)} 
		&= \hat{U}(t,t_0)\ket{\Psi(t_0)}
		\\&= \dfrac{1}{\pi} \int \dd^2\alpha_{2\omega}
			\braket{\alpha}{\Phi}
				\hat{U}(t,t_0) 
					\big[
						\hat{D}_{\omega}(\alpha_{\omega})
						\otimes \hat{D}_{2\omega}(\alpha_{2\omega})
					\big]\ket{\text{g}}\otimes \ket{\boldsymbol{0}}
		= \dfrac{1}{\pi} 
					\int \dd^2\alpha_{2\omega}
				\braket{\alpha_{2\omega}}{\Phi}
						\hat{U}_{\boldsymbol{\alpha}}(t,t_0)
							\ket{\text{g}}\otimes \ket{\boldsymbol{0}},
	\end{aligned}
\end{equation}
where $\ket{\boldsymbol{0}}$ denotes the multimode vacuum state.~In going from the second to the third equality, we have defined $\hat{U}_{\boldsymbol{\alpha}}(t,t_0) \equiv 	[\hat{D}_{\omega}(\alpha_{\omega})\otimes \hat{D}_{2\omega}(\alpha_{2\omega})]^\dagger \hat{U}(t,t_0)[\hat{D}_{\omega}(\alpha_{\omega})\otimes \hat{D}_{2\omega}(\alpha_{2\omega})]$, with this newly defined propagator satisfying the Heisenberg equation
\begin{equation}
	i\hbar \pdv{\hat{U}_{\boldsymbol{\alpha}}(t)}{t}
		= \Big[
				\hat{H}_{\text{at}} 
				+ \mathsf{e} \hat{r}
					\big(
						E_{\omega}(t) + E_{2\omega}(t)
						+\hat{E}(t)
					\big)
			\Big]
					\hat{U}_{\boldsymbol{\alpha}}(t).
\end{equation}

When considering HHG processes in the low-depletion regime~\cite{rivera2022strong,stammer2022theory,stammer2024entanglement}, it is found that $\bra{g}U_{\boldsymbol{\alpha}}(t,t_0)\ket{g}\otimes \ket{\boldsymbol{0}} \simeq \hat{\boldsymbol{D}}\big(\boldsymbol{\chi}(t,t_0)\big)\ket{\boldsymbol{0}}$, where $\hat{\boldsymbol{D}}\big(\boldsymbol{\chi}(t,t_0)\big) \equiv \bigotimes_q\hat{D}_q(\boldsymbol{\chi}(t,t_0))$.~Consequently, at any time $t\geq t_0$, the total quantum optical state, upon projecting the final electronic state onto the ground state (corresponding to HHG events), is given by
\begin{equation}\label{Eq:SM:state:final}
	\ket{\Phi(t)}
		= \dfrac{1}{\mathcal{N}\pi}
				\int \dd^2 \alpha_{2\omega}
					\braket{\alpha_{2\omega}}{\Phi}
						\big[
							\hat{D}_\omega(\alpha_{\omega})
							\otimes \hat{D}_{2\omega}(\alpha_{2\omega})
						\big]
						\hat{\boldsymbol{D}}\big(\boldsymbol{\chi}(t,t_0;\alpha_{2\omega})\big)
						\ket{\boldsymbol{0}},
\end{equation}
where $\mathcal{N}^2 = \abs{\braket{\text{g}}{\Psi(t)}}^2$ is a normalization constant. In what follows, we denote $\boldsymbol{\chi}(t,t_0;\alpha_2\omega) \equiv \boldsymbol{\chi}(\alpha_{2\omega})$.

Let us now consider the evaluation of expectation values of polynomials of normally ordered creation and annihilation operators acting on the harmonic modes $q > 2$, i.e., $\hat{a}^{\dagger n}\hat{a}^n$. For this class of observables, we find
\begin{equation}\label{Eq:SM:expval:n}
	\begin{aligned}
	\expval{\hat{a}^{\dagger n}\hat{a}^n}
		&= \dfrac{1}{\mathcal{N}^2\pi^2}
				\int \dd^2 \alpha_{2\omega}\int \dd^2 \beta_{2\omega}
					\braket{\alpha_{2\omega}}{\Phi}
					\braket{\Phi}{\beta_{2\omega}}
					[\chi_q(\beta_{2\omega})]^{*n}
					[\chi_q(\alpha_{2\omega})]^{n}
					\\&
					\hspace{4cm}\times
					\bra{\boldsymbol{0}}
						\big[
						\hat{D}_\omega(\alpha_{\omega})
						\otimes \hat{D}_{2\omega}(\beta_{2\omega})
						\big]^\dagger		\hat{\boldsymbol{D}}^\dagger\big(\boldsymbol{\chi}(\beta_{2\omega})\big)
						\big[
						\hat{D}_\omega(\alpha_{\omega})
						\otimes \hat{D}_{2\omega}(\alpha_{2\omega})
						\big]	\hat{\boldsymbol{D}}\big(\boldsymbol{\chi}(\alpha_{2\omega})\big)
					\ket{\boldsymbol{0}}
		\\&\equiv
			\dfrac{1}{\mathcal{N}^2\pi^2}
			\int \dd^2 \alpha_{2\omega}\int \dd^2 \beta_{2\omega}
			\braket{\alpha_{2\omega}}{\Phi}
			\braket{\Phi}{\beta_{2\omega}}
			[\chi_q(\beta_{2\omega})]^{*n}
			[\chi_q(\alpha_{2\omega})]^{n}
			w(\boldsymbol{\alpha},\boldsymbol{\beta},\boldsymbol{\chi}),
	\end{aligned}
\end{equation}
where, for simplicity, we have defined
\begin{equation}
	w(\boldsymbol{\alpha},\boldsymbol{\beta},\boldsymbol{\chi})
		\equiv 	
				\bra{\boldsymbol{0}}
			\big[
			\hat{D}_\omega(\alpha_{\omega})
			\otimes \hat{D}_{2\omega}(\beta_{2\omega})
			\big]^\dagger		\hat{\boldsymbol{D}}^\dagger\big(\boldsymbol{\chi}(\beta_{2\omega})\big)
			\big[
			\hat{D}_\omega(\alpha_{\omega})
			\otimes \hat{D}_{2\omega}(\alpha_{2\omega})
			\big]	\hat{\boldsymbol{D}}\big(\boldsymbol{\chi}(\alpha_{2\omega})\big)
			\ket{\boldsymbol{0}}.
\end{equation}
This function is generally complex, and its absolute value satisfies $\abs{w(\boldsymbol{\alpha},\boldsymbol{\beta},\boldsymbol{\chi})} \propto e^{-\abs{\beta - \alpha}^2}e^{-\abs{\boldsymbol{\chi}(\beta_{2\omega})-\boldsymbol{\chi}(\alpha_{2\omega})}^2}$, i.e., it decays exponentially as the phase-space separation between $\alpha_{2\omega}$ and $\beta_{2\omega}$ increases.~For instance, separations of order $\abs{\beta - \alpha} \sim 5$ already suppress $\abs{w(\boldsymbol{\alpha},\boldsymbol{\beta},\boldsymbol{\chi})}$ by three orders of magnitude.~In the present context, such phase space correspond to variations of at most few photons in the driving field, or equivalently to electric field variations of order $E = 2 g \alpha \sim 10^{-7}$ a.u.~By contrast, the strong-field response of the system becomes significantly modified only when variations involve thousands of photons, corresponding to field strengths of order $10^{-2}$ a.u. Consequently, over the region of phase space that effectively contributes to the integrals, the induced displacements satisfy $\boldsymbol{\chi}(\alpha_{2\omega}) \simeq \boldsymbol{\chi}(\beta_{2\omega})$ and the function can be approximated by the coherent state overlap, $w(\boldsymbol{\alpha},\boldsymbol{\beta},\boldsymbol{\chi}) \simeq \braket{\beta_{2\omega}}{\alpha_{2\omega}}$ provided that $\abs{\boldsymbol{\chi}_{q=2}(\alpha_{2\omega})} \lesssim 1$. Under these conditions, which are naturally satisfied in the strong-field regime, Eq.~\eqref{Eq:SM:expval:n} reduces to
\begin{equation}
	\expval{\hat{a}^{\dagger n}\hat{a}^n}
		\simeq \dfrac{1}{\pi}
			\int \dd^2\alpha_{2\omega} 
				\abs{\chi_q(\alpha_{2\omega})}^{2n}
				\int \dd^2 \beta_{2\omega}
						\braket{\alpha_{2\omega}}{\Phi}
						\braket{\Phi}{\beta_{2\omega}}
						\braket{\beta_{2\omega}}{\alpha_{2\omega}},
\end{equation}
where we have taken into account that $\mathcal{N} = 1$. By identifying the resolution of the identity in the coherent state basis, we finally obtain
\begin{equation}
	\expval{\hat{a}^{\dagger n}\hat{a}^n}
		\simeq \dfrac{1}{\pi^2}
			\int \dd^2\alpha_{2\omega} 
				\abs{\chi_q(\alpha_{2\omega})}^{2n}
				\abs{\braket{\alpha_{2\omega}}{\Phi}}^2
		= \int \dd^2\alpha_{2\omega} 
				\abs{\chi_q(\alpha_{2\omega})}^{2n}
				Q_{2\omega}(\alpha_{2\omega}),
\end{equation}
where $Q_{2\omega}(\alpha_{2\omega})\equiv \pi^{-1}\abs{\braket{\alpha_{2\omega}}{\Phi}}^2$ is the Husimi function of the state.~This result coincides with the expressions presented in Sec.~\ref{Sec:App:QO:obs}.

More generally, tracing out the fundamental mode in Eq.~\eqref{Eq:SM:state:final} results incoherent mixtures of coherent states in the harmonic modes, for the same reasons as discussed above: significant modifications in the harmonic response require the exchange of a huge number of photons from the driving field.~This dispair variations, however, result in highly correlated states between the fundamental and the harmonic modes.~As a consequence, when the harmonics are considered individually in the presence of strongly squeezed driving fields, or even when pairs of harmonics are analyzed, their corresponding photon statistics remain comparable with those of classical coherent mixtures.~This can be more explicitly seen by tracing out the fundamental mode (and all modes except the $q$th one) in Eq.~\eqref{Eq:SM:state:final}, which results in
\begin{equation}
    \hat{\rho}_q(t)
        = \tr_{\{q' \neq q\}}\big[\dyad{\Phi(t)}\big]
        =\dfrac{1}{\mathcal{N}^2\pi^2}
			\int \dd^2 \alpha_{2\omega}\int \dd^2 \beta_{2\omega}
			\braket{\alpha_{2\omega}}{\Phi}
			\braket{\Phi}{\beta_{2\omega}}
			\dyad{\chi_q(\alpha_{2\omega})}{\chi_q(\beta_{2\omega})}
			w(\boldsymbol{\alpha},\boldsymbol{\beta},\boldsymbol{\chi}),
\end{equation}
and that under the conditions mentioned above yields
\begin{equation}
    \hat{\rho}_q(t)
        \simeq
        \dfrac{1}{\pi}
			\int \dd^2 \alpha_{2\omega}\ Q_{2\omega}(\alpha_{2\omega})
            \dyad{\chi_q(\alpha_\omega)}.
\end{equation}

This absence of genuinely non-classical features on the harmonic radiation has also been reported in Ref.~\cite{tzur_generation_2024}, where squeezing in the harmonic modes was observed only for displaced squeezed states with $r \sim 1$, whereas for bright squeezed stats $(r \sim 10)$, the harmonics did not exhibit sub-vacuum noise fluctuations.~Finally, this also explains why the use of coherent state expansions, as considered in this subsection, results in the same expressions as those obtained when using the so-called classical limit~\cite{wang_high-order_2025}.
	
We also note that, under the aforementioned conditions, the expectation values to the squared obey the inequality
\begin{equation}\label{Eq:SM:CSI:analysis}
	\langle \hat{a}^{\dagger n}\hat{a}^n\rangle^2
		\leq \dfrac{1}{\pi} \int \dd^2 \alpha_{2\omega} \abs{\chi_q(\alpha_{2\omega})}^{2n} Q_{2\omega}(\alpha),
\end{equation}
which for $n=1$, implies $g^{(2)}(0) \geq 1$.~The situation changes qualitatively when $\abs{\boldsymbol{\chi}_{q=2}(\alpha_{2\omega})} \gg 1$.~In this case, we explicitly have
\begin{equation}
	\begin{aligned}
	\expval{\hat{a}^{\dagger n}\hat{a}^n}
		&\simeq \dfrac{1}{\mathcal{N}^2\pi^2}
			\int \dd^2\alpha_{2\omega} 
				\abs{\chi_q(\alpha_{2\omega})}^{2n}
					\int \dd^2 \beta_{2\omega}
					\braket{\alpha_{2\omega}}{\Phi}
					\braket{\Phi}{\beta_{2\omega}}
					\braket{\beta_{2\omega}+\chi_{2}(\beta_{2\omega})}{\alpha_{2\omega} +\chi_{2}(\alpha_{2\omega})}
		\\&\equiv	\dfrac{1}{\mathcal{N}^2\pi^2}\int \dd^2\alpha_{2\omega} 
				\abs{\chi_q(\alpha_{2\omega})}^{2n}
					G(\alpha_{2\omega}),
	\end{aligned}
\end{equation}
where the function $G(\alpha_{2\omega})$ is generally complex and satisfies the normalization condition $\pi^{-2}\mathcal{N}^{-2}\int \dd^2 \alpha G(\alpha_{2\omega}) = 1$. Since $G(\alpha_{2\omega})$ is no longer a positive definite function, the application of the Cauchy-Schwarz inequality as in Eq.~\eqref{Eq:SM:CSI:analysis} is not valid.~Consequently, no general bound enforcing $g^{(2)}(0) \geq 1$ can be established. 

Physically, this regime corresponds to moderate depletion of the ground state, characterized by $\mathcal{N}^2 = \abs{\braket{\text{g}}{\Psi(t)}}^2< 1$, indicating an increased probability of ionization.~In this case, however, additional quantum correlations between the fundamental and the harmonic modes arise, and entanglement and squeezing effects must be explicitly taken into account, as discussed in Ref.~\cite{stammer2024entanglement}.

\section{Evaluation of the linear entropy}\label{Sec:App:Lin:Entropy}
The linear entropy is not, strictly speaking, a quantum optical observable; therefore, the derivation presented for Eq.~\eqref{Eq:App:Obs:lim} is not formally valid in this case.~The linear entropy is defined through the purity, which, in terms of Eq.~\eqref{Eq:Meth:state:2}, can be expressed as
\begin{equation}
	\begin{aligned}
	\gamma_q = \tr(\hat{\rho}_q^2)
		= \int \dd^2 \alpha_1
				\int \dd^2 \beta_1
					\int \dd^2 \alpha_2
						\int \dd^2 \beta_2&\
							\dfrac{P(\alpha_1,\beta_1^*)}{\braket{N\chi_{\beta_1^*,q}(t)}{N\chi_{\alpha_1,q}(t)}}
							\dfrac{P(\alpha_2,\beta_2^*)}{\braket{N\chi_{\beta_2^*,q}(t)}{N\chi_{\alpha_2,q}(t)}}
							\\ &\times 
							\braket{N\chi_{\beta^*_2,q}}{N\chi_{\alpha_1,q}}
							\braket{N\chi_{\beta^*_1,q}}{N\chi_{\alpha_2,q}},
	\end{aligned}
\end{equation}
where the condition $N\chi_{\alpha,q} = 0$ corresponds to a vacuum state and therefore to a pure state.~Nevertheless, this formulation enables the introduction of both the quasi-thermodynamic and classical limits.~Analogously to Eq.~\eqref{Eq:App:Obs:lim}, the evaluation of the purity for the generated harmonics does not introduce any additional dependence on $N$ or $V$, so both limits can be consistently applied.~This yields
\begin{equation}
	\gamma_q 
	= \int \dd \varepsilon_{\alpha}\! \int \dd \varepsilon_{\alpha}'
			\mathcal{Q}(\varepsilon_{\alpha})
				\mathcal{Q}(\varepsilon_{\alpha}')
					\abs{\braket{\varrho d_{\varepsilon_{\alpha}}(\omega_q)}{\varrho d_{\varepsilon_{\alpha}'}(\omega_q}}^2.
\end{equation}

\section{Numerical sampling of homodyne-like measurements}\label{Sec:App:HT}
\begin{figure}[hb!]
	\centering
	\includegraphics[width=1\textwidth]{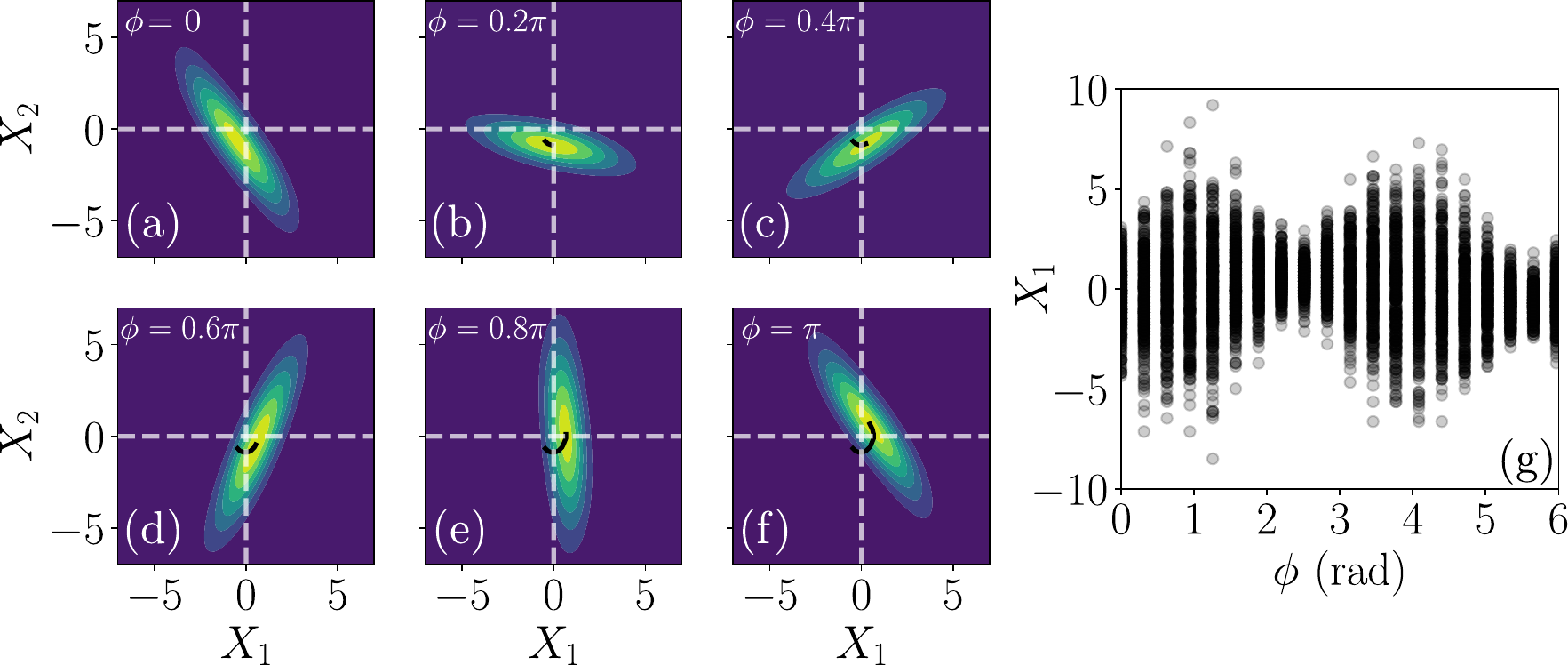}
	\caption{(a)-(f) Wigner functions of the 12th harmonic for different two-color phases $\phi$. The black line traces the rotation of the Wigner function's maximum in phase space as $\phi$ varies. (g) AQT-trace of the 12th harmonic order for the quadrature $\hat{X}_1 = \hat{a}^\dagger + \hat{a}$.}
	\label{fig:Homodyne_backup}
\end{figure}

The main idea behind the \textit{attosecond quantum tomography (AQT)} technique is that, by varying the two-color phase, one can effectively rotate the quantum state in phase space. This is illustrated in Fig.~\ref{fig:Homodyne_backup}~(a)-(f) for the 12th harmonic orders, where a full rotation of the Wigner function is observed as $\phi$ is going from $0$ to $\pi$. A crucial point, however, is that the properties of the quantum state itself depend on the two-color phase $\phi$. This is evident, for example, in the analysis of $g^{(2)}(0)$ and $\Delta X_{\theta}^2$, and is also reflected in Fig.~\ref{fig:Homodyne_backup}~(a)-(f). There, both the amount of squeezing and the position of the Wigner function maxima vary with $\phi$, with the latter oscillating around the origin (black curve). This reveals one of the key caveats of this method: unlike true homodyne detection, where the local oscillator is independent of the state under investigation, here the effective local oscillator and the signal state are intrinsically coupled through the same parameter $\phi$.

In any case, one can still perform a homodyne-like measurement on the selected harmonic mode and construct a corresponding AQT-trace, as illustrated in Fig.~\ref{fig:Homodyne_backup}~(g). To do so, we employ a numerical sampling procedure analogous to that used in Ref.~\cite{rivera-dean_quantum_2025}.~First, the state $\hat{\rho}_q$ is numerically represented in the Fock basis using the \texttt{QuTiP} package in Python~\cite{johansson_qutip_2012,johansson_qutip_2013}, with a cutoff $n_{\text{cutoff}} = 200$, which is sufficient for the states considered here.~We then define the quadrature operator $\hat{X} = \hat{a} + \hat{a}^\dagger$. Its eigenvalues $\{\lambda_i\}$ correspond to the possible measurement outcomes, with its eigenstates $\{\ket{\varphi_i}\}$ onto which $\hat{\rho}_q(\phi)$ is projected.~Consequently, the probabilities of obtaining each outcome are given by $ p(\lambda_i,\phi) = \langle \varphi_i\vert \hat{\rho}_q(\phi)\vert \varphi_i \rangle$.~Knowing both $\{\lambda_{i}\}$ and $\{p(\lambda_i,\phi)\}$, we simulate the homodyne-like outcomes  by sampling numerically from this probability distribution using the \texttt{random} package in Python~\cite{python-random}. This is repeated for each value of $\phi$: in total, we consider 20 phase settings, with 500 shots per setting. Finally, we reconstruct the AQT- distribution [main text Fig.~4~(a),(b)] by applying the inverse Radon transformation~\cite{leonhardt_measuring_1997,lvovsky_continuous-variable_2009}, with integration limits set to $k_c = 3$.

\section{Insights about the semiclassical action and the electronic trajectories}\label{Sec:App:Trajectories}

\subsection{Classical fields}
When a strong coherent state $\omega$ field is combined with a perturbative $2\omega$ coherent component, the total semiclassical action acquires an additional correction due to the $2\omega $ field~\cite{dahlstrom2011quantum}
\begin{equation}
	S(p,t_2,t_1) =  S_{\omega}(p,t_2,t_1) + \sigma(p,t_2,t_1),
\end{equation}
so that the total intensity of the $q$th harmonic order can be expressed as~\cite{pedatzur2015attosecond}
\begin{equation}
	I_q = I_{0,q}\left\{
	\begin{aligned}
		&\abs{\cos(\sigma)}^2 \ \text{if $q$ is odd},
		\\&\abs{\sin(\sigma)}^2 \ \text{if $q$ is even},
	\end{aligned}
	\right.
\end{equation}
with this expression being obtained via a saddle-point analysis. Here, we assume $I_{0,q} = I_0$ independent of the harmonic order. While this is not strictly correct---since saddle-points vary with harmonic order---for consecutive harmonics order this approximation is reasonable~\cite{pedatzur2015attosecond}.

In the saddle-point method, $\sigma \in \mathbbm{C}$, which we write as $\sigma = \sigma_x + i \sigma_y$. Then
\begin{equation}
	\begin{aligned}
		&\abs{\cos\sigma}^2
		= \cos[2](\sigma_x)\cosh[2](\sigma_y) + \sin[2](\sigma_x)\sinh[2](\sigma_y)
		\\&\abs{\sin\sigma}^2
		= \sin[2](\sigma_x)\cosh[2](\sigma_y) + \cos[2](\sigma_x)\sinh[2](\sigma_y),
	\end{aligned}
\end{equation}
so that, when taking the difference and sum between two consecutive harmonic orders, we obtain the approximate relation
\begin{equation}
	\begin{aligned}
		&I_{\text{odd}}-I_{\text{even}}
		= I_0 
		\big[
		\cos[2](\sigma_x) - \sin[2](\sigma_x)
		\big]
		= I_0 \cos(2\sigma_x)	,
		\\&
		I_{\text{odd}}+I_{\text{even}}
		= I_0 
		\big[
		\cosh[2](\sigma_y) + \sinh[2](\sigma_y)
		\big]
		=I_0 \cosh(2\sigma_y),
	\end{aligned}
\end{equation}
which results in~\cite{pedatzur2015attosecond}
\begin{equation}\label{Eq:inversion}
	\sigma_x =\dfrac12 \cos[-1](\dfrac{I_{\text{odd}}-I_{\text{even}}}{I_0}),
	\quad 
	\sigma_y =\dfrac12 \cosh[-1](\dfrac{I_{\text{odd}}+I_{\text{even}}}{I_0}).
\end{equation}

This result is of particular importance: it shows that information about the electron dynamics can be extracted directly from the harmonic spectrum.

\subsection{Squeezed driving fields}
In contrast, when considering squeezed driving fields, the intensity of the $q$th harmonic order is given by
\begin{equation}
	I_q
	= \int \dd \varepsilon_\alpha
	\mathcal{Q}(\varepsilon_\alpha)
	I_0(\varepsilon_\alpha)
	\bigg[
	\delta_{q,\text{odd}}\abs{\cos(\sigma)}^2
	+ \delta_{q,\text{even}}\abs{\sin(\sigma)}^2
	\bigg],
\end{equation}
and evaluating the difference between two consecutive harmonics yields
\begin{equation}
	I_{\text{odd}}-I_{\text{even}}
	\approx \int \dd \varepsilon_\alpha \mathcal{Q}(\varepsilon_\alpha)I_0(\varepsilon_\alpha) \cos[2](\sigma_r),
\end{equation}
from which it is clear that inversions of the type performed in Eq.~\eqref{Eq:inversion} are no longer straightforward, as they were in the classical scenario. Thus, with squeezed light, extracting information about the value of $\sigma$ is significantly more involved. In this case, $\sigma$ is determined from semiclassical saddle-point equations, i.e., for each value of $\alpha$ independently.

An alternative approach can be obtained through homodyne measurements of the generated harmonic orders. While there may exist simpler methods from an experimental perspective, here we proceed to justify this strategy, which essentially lies on the possibility of using it for extracting amplitude and phase of the harmonic radiation. The outcome of an homodyne measurement, where $\theta$ denotes the phase of the local oscillator, is generally given by
\begin{equation}
	\begin{aligned}
		\langle X_{\theta}(\omega) \rangle
		&\propto 
		\int \dd\alpha \mathcal{Q}(\varepsilon_\alpha)
		\bigg[
		\langle d_{\varepsilon_\alpha}(\omega)\rangle e^{i\theta}
		+ \langle d_{\varepsilon_\alpha}(\omega)\rangle^* e^{i\theta}
		\bigg]
		\\&\equiv
		\int \dd\varepsilon_\alpha \mathcal{Q}(\varepsilon_\alpha)
		\bigg[
		\text{Re}[\langle d_{\varepsilon_\alpha}(\omega)\rangle] \cos(\theta)
		+ \text{Im}[\langle d_{\varepsilon_\alpha}(\omega)\rangle] \sin(\theta)
		\bigg],
	\end{aligned}
\end{equation}
which shows that, by suitably choosing $\theta$, one can access the real and imaginary parts of the Fourier transform of the time-dependent dipole moment. For even harmonic orders, this becomes
\begin{equation}
	\langle d_{\varepsilon_\alpha}(\omega)\rangle
	= \abs{x(\omega,\varepsilon_\alpha)}^2 e^{i\arg(x)} \sin(\sigma_{\varepsilon_\alpha}),
\end{equation}
so that the homodyne signal for even harmonics can be expressed as
\begin{equation}
	\langle X_{\theta}(\omega) \rangle
	\approx \int \dd \varepsilon_\alpha
	\mathcal{Q}(\varepsilon_\alpha)
	\abs{x(\omega,\varepsilon_\alpha)}^2
	\Big[
	\sigma_x(\varepsilon_\alpha)
	\cos(\theta +\arg(x))
	+ \sigma_y(\varepsilon_\alpha)
	\sin(\theta + \arg(x))
	\Big],
\end{equation}
which holds whenever $\sin(\sigma_{\varepsilon_\alpha}) \approx \sigma_{\varepsilon_\alpha}$, a condition expected to be valid in the perturbative squeezing regime. By appropriately varying $\theta$, one can therefore directly probe the expectation values of the real and imaginary parts of $\sigma$ with respect to the Husimi function. The remaining challenge is to determine a suitable method for extracting the phase $\arg(x)$, which would allow for disentangling $\sigma_x$ and $\sigma_y$. In principle, however, all values of $\theta$ can be probed.

\subsection{Gabor transform}
In the context of HHG, the Gabor transform is defined as~\cite{risoud_quantitative_2013}
\begin{equation}
	G(\omega,t)
	= \int \dd t\ d(t) w(t-\tau)e^{-i\omega t},
\end{equation}
where $w(t)$ is a window function, chosen here to be a Gaussian of width $\delta$
\begin{equation}
	w(t) = \dfrac{1}{\delta \sqrt{\pi}} \exp[-\dfrac{t^2}{\delta^2}].
\end{equation}

The Gabor transform therefore acts as a localized bandpass filter applied to the signal's full spectrum, with the temporal window providing time localization. In this way, it quantifies the likelihood fo emitting a given harmonic frequency around specific instants of time. However, due to the uncertainty principle, frequency and time of emission cannot be accessed simultaneously with perfect precision. The width of the window function must thus be carefully adjusted to balance time and frequency resolution.~Here, we set $\delta = 6$ a.u.~\cite{risoud_quantitative_2013}, corresponding to 145 as, while the total signal has a period $T\simeq2.67$ fs. Furthermore, the time-dependent dipole moment is generally given by~\cite{even_tzur_photon-statistics_2023,rivera2025structured}
\begin{equation}
	d(t)
	= \int \dd \varepsilon_\alpha\ \mathcal{Q}(\varepsilon_\alpha)
	\langle \psi_{\varepsilon_\alpha}(t)\vert \hat{d}\vert \psi_{\varepsilon_\alpha}\rangle,
\end{equation}
where $\ket{\psi_{\varepsilon_\alpha}(t)}$ denotes the electronic state evolved under the classical field $E_{\text{cl}}(t) = \text{tr}[\hat{E}(t) \dyad{\alpha_{\omega},\alpha,\{0\}_q}]$. 

\begin{figure}
	\centering
	\includegraphics[width=1\textwidth]{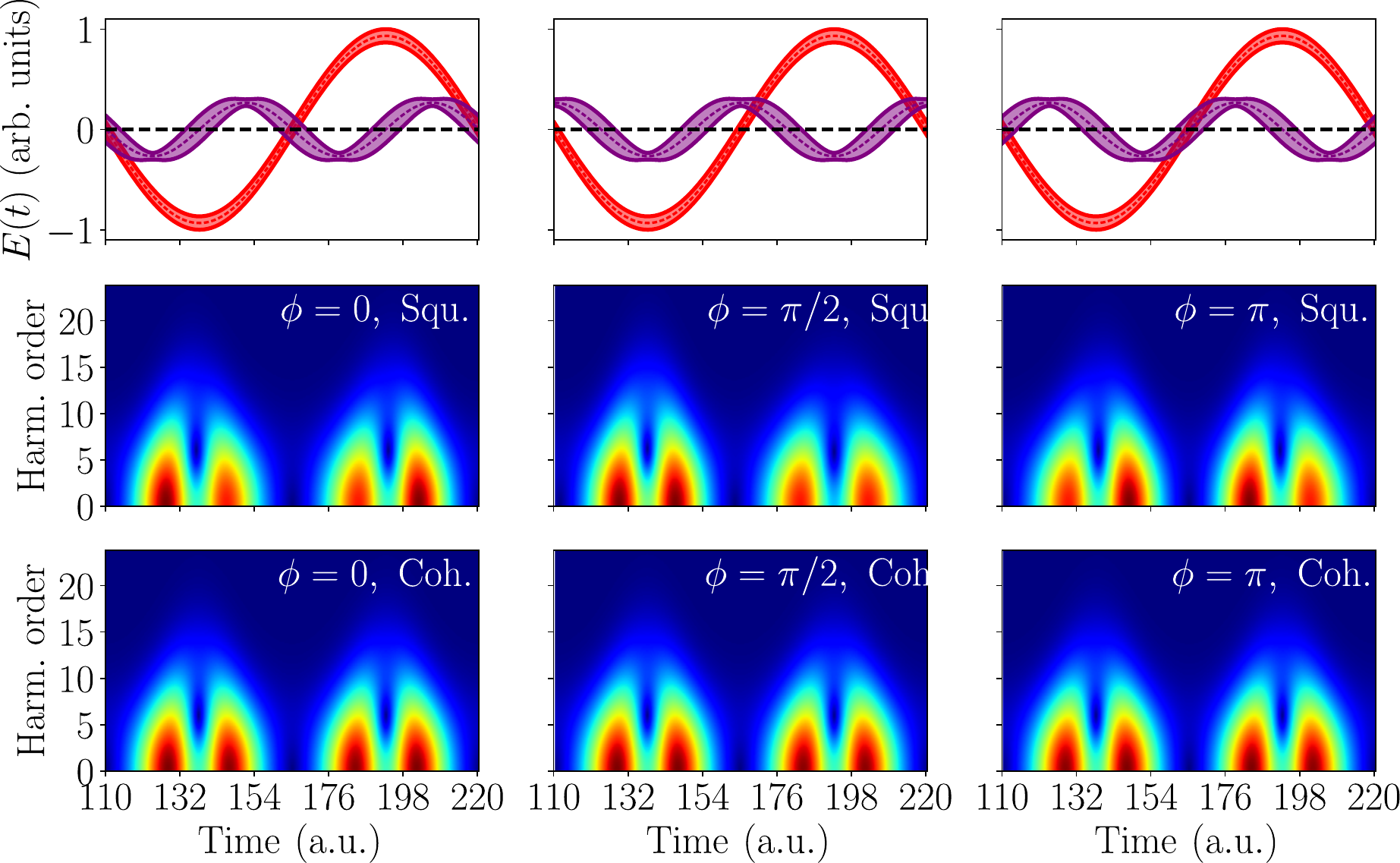}
	\caption{Gabor transform for the case of amplitude squeezing. The first row displays the field configuration, the second the Gabor transform when adding squeezing to the $2\omega$ field ($I_{\text{squ}} = 10^{-6}$ a.u.), and the third row when just having a coherent state.}
	\label{Fig:Gabor:tranf}
\end{figure}

Figure~\ref{Fig:Gabor:tranf} shows the Gabor transform for various values of $\phi$, comparing the case of a squeezed $2\omega$ field (second row) with that of a coherent state. As observed, the Gabor transform is not significantly perturbed by varying the two-color phase when the $2\omega$ field is coherent. This is expected, since it is generally observed that the saddle-point equations are only weakly affected in this case ($I^{(\text{coh})}_{2\omega} = 10^{-2} I^{(\text{coh})}_{\omega}$)~\cite{dahlstrom2011quantum,pedatzur2015attosecond}. In contrast, the presence of squeezing can substantially enhance or suppress the harmonic emission~\cite{stammer_weak_2025}, depending on whether the field fluctuations interfere constructively or destructively with the contribution of the $\omega$ field. Since this interference can be controlled via the two-color delay, squeezing provides an ultrafast means of nonperturbatively controlling the emission time of the harmonic orders with attosecond precision. One may even speculate that, if the amount of squeezing were increased to a nonperturbative level, the emission of harmonic radiation could be confined to specific sub-cycles of the driving field.

\end{document}